\documentclass[twocolumn,tighten,times]{aastex7}
\usepackage{amsmath}
\usepackage{float}
\usepackage{graphicx}
\usepackage{amssymb}
\usepackage{amsbsy}
\usepackage{amsmath}
\usepackage{amsfonts}
\usepackage{mathrsfs}
\usepackage{color}
\usepackage{multirow}
\usepackage{booktabs}	
\usepackage{diagbox}
\usepackage{booktabs}              
\usepackage{makecell}      
\usepackage{array}         

\usepackage{bm}
\DeclareMathAlphabet{\mathbi}{OT1}{ptm}{bx}{it}
\SetMathAlphabet\mathbi{bold}{OT1}{ptm}{bx}{it}

\shorttitle{Accretion Variations Regulate Type Transition}
\shortauthors{Shen et al.}

\begin{document}

\title{\bf\large Accretion-Regulated Type Transitions in Changing-Look AGNs: Evidence from Two-Epoch Spectral Analysis} 

\author{Yu-Heng Shen}
\affiliation{Yunnan Observatories, Chinese Academy of Sciences, Kunming 650011, China}
\affiliation{University of Chinese Academy of Sciences, Beijing 100049, China}
\affiliation{Key Laboratory for the Structure and Evolution of Celestial Objects, Chinese Academy of Sciences, Kunming 650011, China}
\email{shenyuheng@ynao.ac.cn}

\author[0000-0002-2310-0982]{Kai-Xing Lu}
\affiliation{Yunnan Observatories, Chinese Academy of Sciences, Kunming 650011, China} 
\affiliation{University of Chinese Academy of Sciences, Beijing 100049, China} 
\affil{Key Laboratory for the Structure and Evolution of Celestial Objects, Chinese Academy of Sciences, Kunming 650011, China}
\email[show]{lukx@ynao.ac.cn}

\author[0000-0001-9457-0589]{Wei-Jian Guo}
\affiliation{National Astronomical Observatories, Chinese Academy of Sciences, 20A Datun Road, Chaoyang District, Beijing 100101, China}
\email[show]{guowj@bao.ac.cn} 

\author[0000-0003-3823-3419]{Sha-Sha Li}
\affiliation{Yunnan Observatories, Chinese Academy of Sciences, Kunming 650011, China} 
\affiliation{Key Laboratory for the Structure and Evolution of Celestial Objects, Chinese Academy of Sciences, Kunming 650011, China} 
\email{lishasha@ynao.ac.cn}

\author[0000-0002-1530-2680]{Hai-Cheng Feng}
\affiliation{Yunnan Observatories, Chinese Academy of Sciences, Kunming 650011, China} 
\affiliation{Key Laboratory for the Structure and Evolution of Celestial Objects, Chinese Academy of Sciences, Kunming 650011, China}
\email{hcfeng@ynao.ac.cn}

\author{Zhang Yue}
\affiliation{Yunnan Observatories, Chinese Academy of Sciences, Kunming 650011, China} 
\affiliation{University of Chinese Academy of Sciences, Beijing 100049, China} 
\affiliation{Key Laboratory for the Structure and Evolution of Celestial Objects, Chinese Academy of Sciences, Kunming 650011, China}
\email{1398611241@qq.com}

\author{Wen-Zhe Xi}
\affiliation{Yunnan Observatories, Chinese Academy of Sciences, Kunming 650011, China} 
\affiliation{University of Chinese Academy of Sciences, Beijing 100049, China} 
\affiliation{Key Laboratory for the Structure and Evolution of Celestial Objects, Chinese Academy of Sciences, Kunming 650011, China}
\email{xiwenzhe@ynao.ac.cn}

\author{Jian-Guo Wang}
\affiliation{Yunnan Observatories, Chinese Academy of Sciences, Kunming 650011, China}  
\affil{Key Laboratory for the Structure and Evolution of Celestial Objects, Chinese Academy of Sciences, Kunming 650011, China}
\email{wangjg@ynao.ac.cn}

\author{Jin-Ming Bai}
\affiliation{Yunnan Observatories, Chinese Academy of Sciences, Kunming 650011, China} 
\affiliation{University of Chinese Academy of Sciences, Beijing 100049, China} 
\affil{Key Laboratory for the Structure and Evolution of Celestial Objects, Chinese Academy of Sciences, Kunming 650011, China}
\email{baijinming@ynao.ac.cn}

\begin{abstract}
The changing-look active galactic nucleus (CL-AGN), an extraordinary subpopulation of supermassive black holes, 
has attracted growing attention for understanding its nature.
We present an analysis of the spectral properties of 203 low-redshift CL-AGNs ($z<0.35$) 
using two-epoch spectra from SDSS DR16 and DESI DR1 with time baseline ranging from $\sim$1000 to 8000 days, 
based on spectral fitting and decomposition. 
The sample consists of 11.3\% Type 1.0, 26.6\% Type 1.2, 43.1\% Type 1.5, and 19\% Type 1.8/2.0 AGNs. 
The total sample is divided into two datasets: 
Dataset A (110 objects) with minor spectral type variations, likely general AGN variability, 
and Dataset B (93 objects) showing significant type transitions and characteristic turn-on or turn-off behavior.
Our results reveal clear optical continuum and emission-line variability, showing both “bluer-when-brighter” and “redder-when-brighter” trends. 
A strong correlation between the broad H$\beta$/[O~{\sc iii}] ratio and broad H$\alpha$ luminosity ($L_{\rm H\alpha}$), 
${\rm log(H\beta/[O~III])}=(0.63\pm 0.07){\rm log}(L_{\rm H\alpha})-(26.49\pm2.96)\pm0.48$ for Dataset B, 
as well as the correlation between H$\beta$/[O~{\sc iii}] and Eddington ratio ($L_{\rm bol}/L_{\rm Edd}$), 
${\rm log(H\beta/[O~III])}=(0.59\pm 0.08){\rm log}(L_{\rm bol}/L_{\rm Edd})+(1.02\pm0.15)\pm0.53$ for Dataset B, 
suggests that accretion rate variations drive changes in ionizing flux within the broad-line region, 
thereby triggering AGN type transitions. 
These findings underscore the critical role of supermassive black hole accretion processes in refining the AGN unification model. 
Future work should investigate potential connections between stellar evolution in outer accretion disk and the observed scatter in these correlations. 
\end{abstract}

\keywords{\uat{Accretion}{14} --- \uat{Active galactic nuclei}{16} --- \uat{Active galaxies}{17} --- \uat{Galaxy evolution}{594} 
--- \uat{Spectroscopy}{1558} --- \uat{Time domain astronomy}{2109}}

\section{Introduction} \label{intro}  
Active galactic nuclei (AGN) are powered by accretion onto supermassive black holes (SMBHs). 
According to the unified model (\citealt{Antonucci1993,Urry1995}), AGNs exhibit a common layered structure, 
including a dusty toroidal structure (the torus) surrounding an accreting SMBH. 
The broad-line region (BLR), located near the central region and within the inner radius of the dusty torus, emits broad emission lines. 
The narrow-line region (NLR), extending to distances of a few kilo-parsecs, generates narrow emission lines.  

Based on the flux ratio of broad H$\beta$ to [O~{\sc iii}]~$\lambda$5007 emission lines, 
AGNs are categorized into distinct types (\citealt{Osterbrock1977,Winkler1992}): 
Type 1.0 (H$\beta$/[O {\sc iii}]$>$5), 
Type 1.2 (5$>$H$\beta$/[O {\sc iii}]$>$2), 
Type 1.5 (2$>$H$\beta$/[O {\sc iii}]$>$0.33), 
Type 1.8 (0.33$>$H$\beta$/[O {\sc iii}]$>$0), 
Type 2.0 (No broad emission lines detected). 
These classifications arise from variations in viewing angles as described by the unification model. 
On the other hand, a Type 1.0 AGN may pass through some of these subtypes as its ionizing luminosity fluctuates from high to low states, 
which is predicted by photoionization calculation of \cite{Korista2004}. 
These results highlight the inherent diversity among AGN populations (here we referred to as AGN diversity). 

Recently, hundreds of changing-look AGNs (CL-AGN), characterized by the emergence (turn on) or 
disappearance (turn off) of broad emission lines accompanied by extreme continuum flux variability, 
have been detected (e.g., \citealt{MacLeod2016,Yang2018,Guo2019,Graham2020,Zeltyn2024,Guo2024a,Guo2024b}). 
Only a limited number of CL-AGN exhibiting recurring changing-look phenomena accompanied with 
transitions between different AGN types (e.g., Type 1 to Type 1.2/1.5, Type 1.8/1.9, or Type 2, and vice versa) 
have been studied in detail (\citealt{Wang2024,Komossa2024}), such as Mrk~1018 (\citealt{Cohen1986,McElroy2016,Lyu2021,Lu2025,Saha2025}), 
Mrk~590 (\citealt{Denney2014}), NGC~4151 (\citealt{Shapovalova2010,Chen2023}), NGC~2617 (\citealt{Moran1996,Shappee2014,Feng2021}), 
and NGC~1566 (\citealt{Oknyansky2019,Xu2024}). 
Notably, \cite{Lu2025} documented a perfect case where Mrk~1018 underwent full-cycle type transitions, 
covering Types 1.0, 1.2, 1.5, 1.8, and 2.0 over a period of 45 years. 
This case poses significant challenges to the traditional scenarios proposed by the unification model. 

Significant efforts have been made to uncover the mechanisms behind these changes 
(e.g., \citealt{Husemann2016,Krumpe2017,Kim2018,Sniegowska2020,Guo2020,Liu2021,Ricci2023,Wu2023,Veronese2024,Panda2024}). 
Briefly, three scenarios can explain the changing-look behavior: (1) changes in obscuration along the line of sight (e.g., \citealt{Mereghetti2021}), 
(2) extreme variations in the accretion rate due to unresolved feeding mechanisms (e.g., \citealt{Sheng2017,Panda2024,Zeltyn2024,Saha2025}), 
and (3) tidal disruption events (e.g., \citealt{Li2022}). 
Many studies have attributed the changing-look phenomenon in AGNs to extreme variations in the black hole accretion rate 
(e.g., \citealt{Sheng2017,Noda2018,Lyu2021,Veronese2024,Ma2025}). 
Definitely, \cite{Lu2025} found that the type transition in Mrk~1018 is modulated by the accretion rate, 
but the universality of this phenomenon remains unclear. 

In addition, the changing-look process of Mrk~1018 was accompanied by an extreme change in the accretion rate, 
with the Eddington ratio increasing from 2.0$\times10^{-5}$ to 0.02. 
\cite{Lu2025} further found that (1) the broad-line Balmer decrement, defined as the flux ratio of broad Balmer lines (H$\alpha$/H$\beta$), 
initially rises but then declines as the Eddington ratio increases; 
and (2) the shape of the broad-line profile, parameterized by the line-width ratio of FWHM to line dispersion (i.e., FWHM/$\sigma_{\rm line}$), 
may depend on the structure of the accretion disk. We will strive to verify these properties in a larger CL-AGN sample. 
In this study, we use the selected CL-AGN sample with redshifts less than 0.35 and their optical spectra, 
which cover the H$\alpha$ and H$\beta$ emission lines, to investigate the universality of these phenomenon newly discovered in Mrk~1018. 
Throughout the paper, we use a cosmology with $H_{0}$=67~km~s$^{-1}$~Mpc$^{-1}$, $\Omega_{M}$=0.32 (\citealt{Planck2020}). 

\begin{figure*}[ht!]
\includegraphics[angle=0,width=0.999\textwidth]{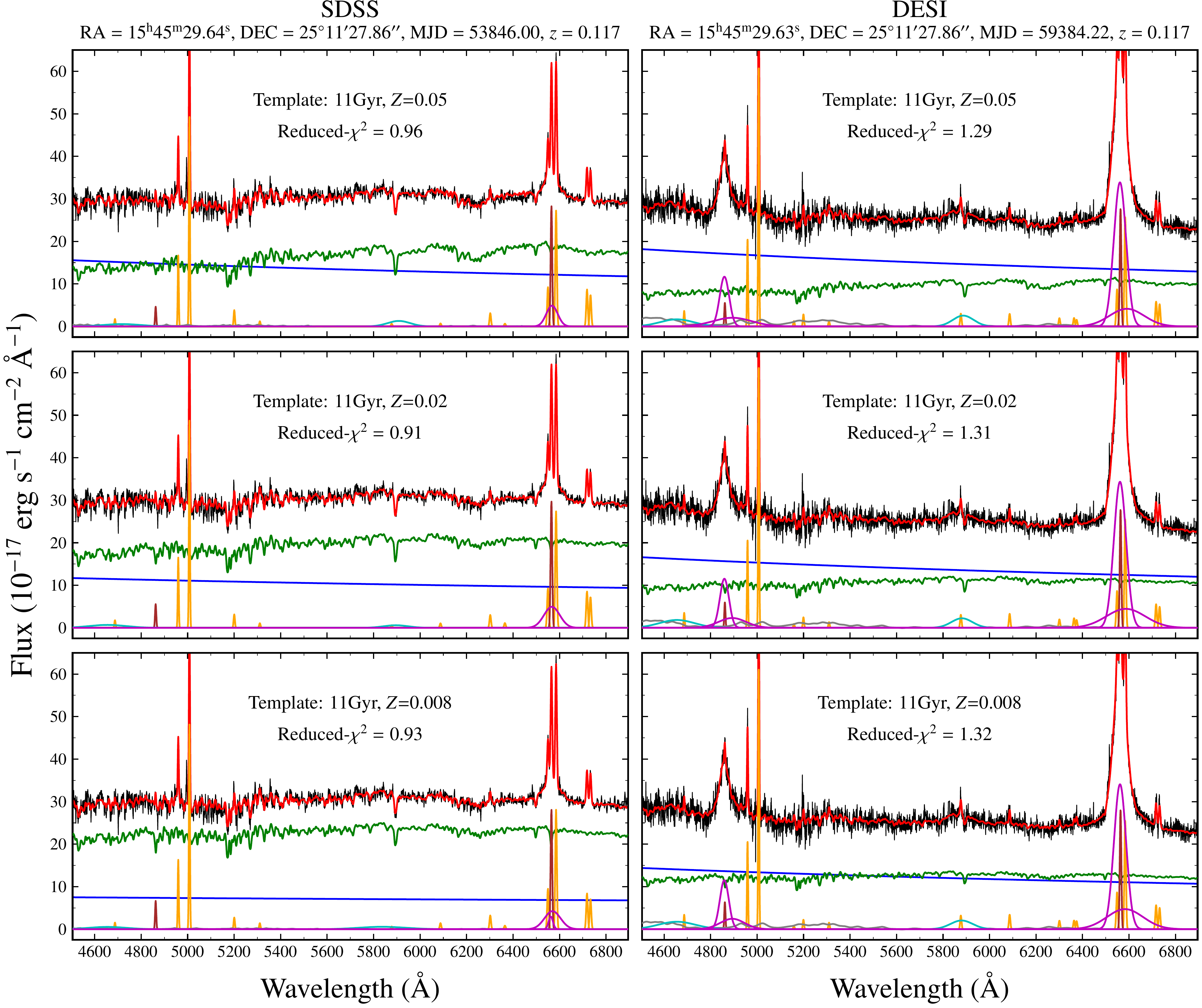}
\caption{
An example of spectral fitting and decomposition of the SDSS (left) and DESI (right) spectra 
for the changing-look AGN~J154529.63+251127.86 ($z$=0.117). The host-galaxy starlight of each epoch spectrum 
was fitted by three host-galaxy templates, they with stellar population age of 11 Gyr and metallicities of 0.008, 
0.02, and 0.05 (more considerations refer to Section~\ref{specfit}), are marked in each panel, respectively. 
Each rest-frame spectrum (black) was fitted over the rest-frame wavelength range 4500–6900~\AA, 
the total model is shown in red. Fitting components include the AGN continuum (blue), iron multiplets (gray), 
host galaxy (green), broad Balmer lines (magenta), narrow Balmer lines (brown), 
broad helium lines (cyan), and other narrow lines including helium lines, [O~{\sc iii}], [N~{\sc ii}], and [S~{\sc ii}] (orange). 
}
\label{fig:specfit}
\end{figure*}

\section{Sample and Spectral fitting}
\subsection{Sample}
\cite{Guo2025} constructed a sample of 561 CL-AGNs at redshifts $z<0.9$ 
by combining the first data release from the Dark Energy Spectroscopic Instrument (DESI DR1) 
with the 16th data release from the Sloan Digital Sky Survey (SDSS DR16). 
The sample selection employed a multi-step approach involving [O~{\sc iii}]-based spectral flux calibration, 
pseudophotometry measurements, and rigorous visual inspection (see Figure 4 of \citealt{Guo2025}).

For the present study, we extracted a subsample of CL-AGNs with redshifts $z<0.35$ from the \cite{Guo2025} catalog. 
This redshift threshold was specifically chosen to ensure that both the broad H$\alpha$ and H$\beta$ emission lines 
fall within the spectral coverage of both SDSS and DESI instruments, 
enabling detailed analysis of these critical Balmer lines across different observational epochs. 
The resulting sample comprises 203 CL-AGNs, including the well-studied Mrk~1018. 

Our selection aims to investigate the emission line properties of CL-AGNs and 
explore newly identified phenomena in Mrk~1018, 
particularly accretion-regulated type transitions (see Section~\ref{intro} and \citealt{Lu2025}). 
For each object in this sample, we compiled optical spectra from two distinct epochs: 
one from SDSS DR16 and another from DESI DR1. 
Columns (1)-(5) of Table~\ref{tab:para} provide the source identification, 
coordinates (RA, DEC), redshift ($z$), and Modified Julian Date (MJD) for each epoch's spectrum, respectively. 

The observed time baselines between the two spectral epochs in this sample span approximately 
1000 to 8000 days (see Table~\ref{tab:stat} for detailed statistical results). 
Compared with previous studies on CL-AGN samples, this coverage aligns well with established works.  
For instance, \cite{Graham2020} implemented explicit source filtering based on a minimum temporal separation 
of 500 days between spectral epochs to ensure reliable variability detection. 
Similarly, \cite{Zeltyn2024} recently identified 10 new CL-AGNs from SDSS-V first-year data, 
with historical epochs spanning approximately 7 to 20 years prior to follow-up observations.
Notably, \cite{Panda2024} conducted a comprehensive investigation of 93 CL-AGNs with repeat SDSS spectroscopy, 
reporting temporal baselines ranging from $\sim$1000 to 6000 days, 
sufficient to detect significant changes in both optical luminosity and spectral slope. 
Our intervals ($\sim$1000 to 8000 days) match these ranges, 
offering enhanced temporal leverage to capture characteristic CL-AGN variability and spectral transitions. 
This sample provides an opportunity to study the physical mechanisms driving spectral state transitions in AGNs. 
Further details on the sample properties are provided in Section~\ref{sprop}. 

\subsection{Spectral fitting} \label{specfit}  
Following previous studies, we conducted spectral fitting and decomposition to separate the blended components in the spectral measurement of AGNs 
(e.g., \citealt{Hu2008,Dong2008,Shen2011,Stern2013,Sun2018,Guo2018,Guo2019,Lu2021,Lu2022}). 
Our spectral fitting includes the following models/components: 
(1) A power-law ($f_{\lambda}\propto \lambda^{\alpha}$, $\alpha$ is the spectral index) represents the AGN continuum. 
(2) The Fe~{\sc ii} template from \cite{Boroson1992} models the Fe multiplets. 
(3) A stellar template from \cite{Bruzual2003} and \cite{Lu2006} accounts for host-galaxy starlights (more details refer to below). 
(4) One to three Gaussians are used for each broad hydrogen Balmer lines (i.e., broad H$\alpha$ and H$\beta$ lines). 
(5) Two Gaussians model [O~{\sc iii}] $\lambda$4959 and [O~{\sc iii}] $\lambda$5007 with a fixed flux ratio of $1/3$. 
(6) Single Gaussian are adopted for each other components, including the broad and narrow Helium lines, the narrow Balmer lines, 
[N~{\sc ii}] $\lambda$6548 and [N~{\sc ii}] $\lambda$6583 with a flux ratio of $1/2.96$ (\citealt{Dong2008}), 
and [S~{\sc ii}] $\lambda$6716 and [S~{\sc ii}] $\lambda$6731. 
Additionally, several narrow forbidden lines are modeled using single Gaussian. 

After correcting the Galactic extinction using \cite{Schlegel1998}'s extinction map and accounting for redshift, 
we simultaneously fitted above models to each corrected spectrum in rest-frame wavelength range of 4500~\AA~to 6900~\AA. 
To reduce the degrees of freedom, 
the broad H$\gamma$ line was excluded; 
all narrow emission lines share identical velocity and shift for each spectrum. 
Spectral fitting was performed using the DASpec software package
\footnote{\url{https://github.com/PuDu-Astro/DASpec}} 
developed by \cite{Du2018a}, which employs the Levenberg-Marquardt algorithm for $\chi^{2}$ minimization. 

We initially employed a set of stellar templates to fit each spectrum, 
including stellar population ages of 640 Myr, 900 Myr, 1.4 Gyr, 2.5 Gyr, 5 Gyr, and 11 Gyr, as adopted by \cite{Guo2025}, 
and metallicities of 0.008, 0.02 and 0.05 from \cite{Bruzual2003}, 
as well as six nonnegative independent components (ICs: IC1, IC2, IC3, IC4, IC5, and IC6) 
from \cite{Lu2006}. The ICs were constructed using ensemble learning for independent component analysis (EL-ICA), 
which compresses the synthetic galaxy spectral library of \cite{Bruzual2003} into six nonnegative independent components (see \citealt{Lu2006} for details). 
However, we found that the stellar templates with ages of 640 Myr, 900 Myr, 1.4 Gyr, 2.5 Gyr, and 5 Gyr, along with all ICs, 
failed to adequately fit the host-galaxy components. 
This conclusion is based on three criteria: 
(1) Reduced-$\chi^{2}$ values exceeding 1.5;  
(2) in spectra with very weak broad Balmer lines, 
an unusually strong iron multiplet was required to improve the fit, which is not justified;  
and (3) visually inconsistent proportions between the power-law and host-galaxy components, 
specifically, in spectra lacking clear AGN features such as broad emission lines, 
the power-law component dominated over the host galaxy, which is implausible. 
These issues often occurred simultaneously. 

In contrast, the 11 Gyr stellar templates with metallicities of 0.008, 0.02, and 0.05 yielded acceptable fits across all spectra, 
producing nearly identical reduced-$\chi^{2}$ values (less than 1.5), making it difficult to objectively identify the optimal template. 
As illustrated in Figure~\ref{fig:specfit}, these fits appear reasonable at first glance. 
However, a notable discrepancy arises due to the strong degeneracy between the host-galaxy component and the AGN power-law continuum. 
Specifically, the resulting reduced-$\chi^{2}$ values are similar for each spectrum , 
but the contributions of two components differ visually (see Figure~\ref{fig:specfit}). 
This discrepancy introduces uncertainty into derived spectral parameters. 
To account for this degeneracy, we retain all fitting results in subsequent measurements of spectral properties (see Section~\ref{pm}). 
These strategies differ from that of \cite{Guo2025} in constructing the CL-AGN catalog. 
Additional improvements or differences include two aspects. 
(1) instead of fixing the broad hydrogen Balmer lines with two double Gaussians, 
we use one to three Gaussians to minimize the reduced-$\chi^{2}$. 
Specifically, we fit the spectrum of each target at every epoch using one to three Gaussians, 
with the goal of minimizing the reduced-$\chi^2$. 
(2) given the weakness of the narrow H$\beta$ line, we model it with a single Gaussian, 
in contrast to the two-Gaussian approach adopted by \cite{Guo2025}. 

\begin{figure*}[ht!]
\centering
\includegraphics[angle=0,width=0.99\textwidth]{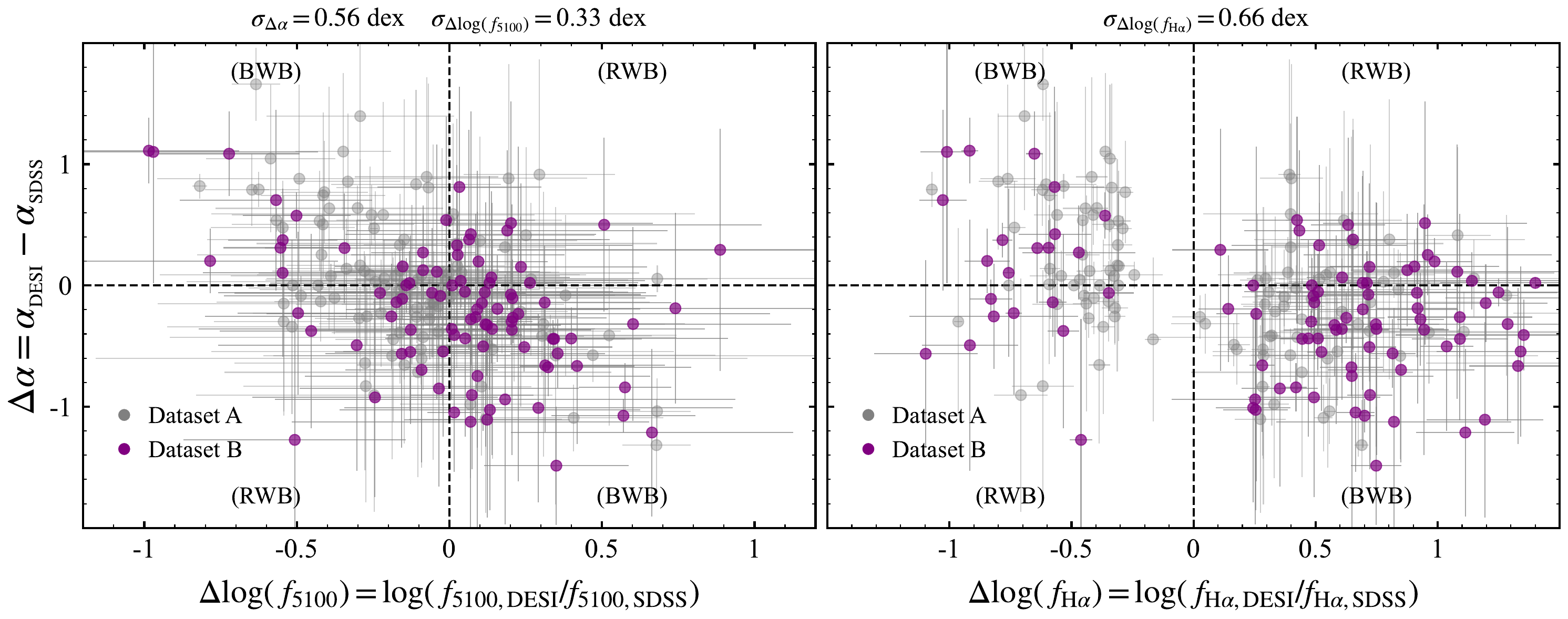}
\caption{
The variation analysis in the spectral index ($\alpha$), optical continuum flux at 5100~\AA\ ($f_{\rm 5100}$), 
and broad H$\alpha$ flux for the Total sample, Dataset A, and Dataset B. 
The standard deviations of these variations, including spectral index ($\Delta \alpha$), 
optical continuum flux (log($f_{\rm 5100,DESI}/f_{\rm 5100,SDSS}$), in magnitudes), 
and broad H$\alpha$ flux (log($f_{\rm H\alpha,DESI}/f_{\rm H\alpha,SDSS}$)), 
are labeled in the titles of the two panels for the Total sample. 
By examining the variation relations between log($f_{\rm 5100,DESI}/f_{\rm 5100,SDSS}$) and $\Delta \alpha$, 
and between log($f_{\rm H\alpha,DESI}/f_{\rm H\alpha,SDSS}$) and $\Delta \alpha$, 
both “bluer-when-brighter (BWB)” and “redder-when-brighter (RWB)” trends are found and marked in the figure.} 
\label{pl}
\end{figure*}

\begin{figure}[ht!]
\centering
\includegraphics[angle=0,width=0.47\textwidth]{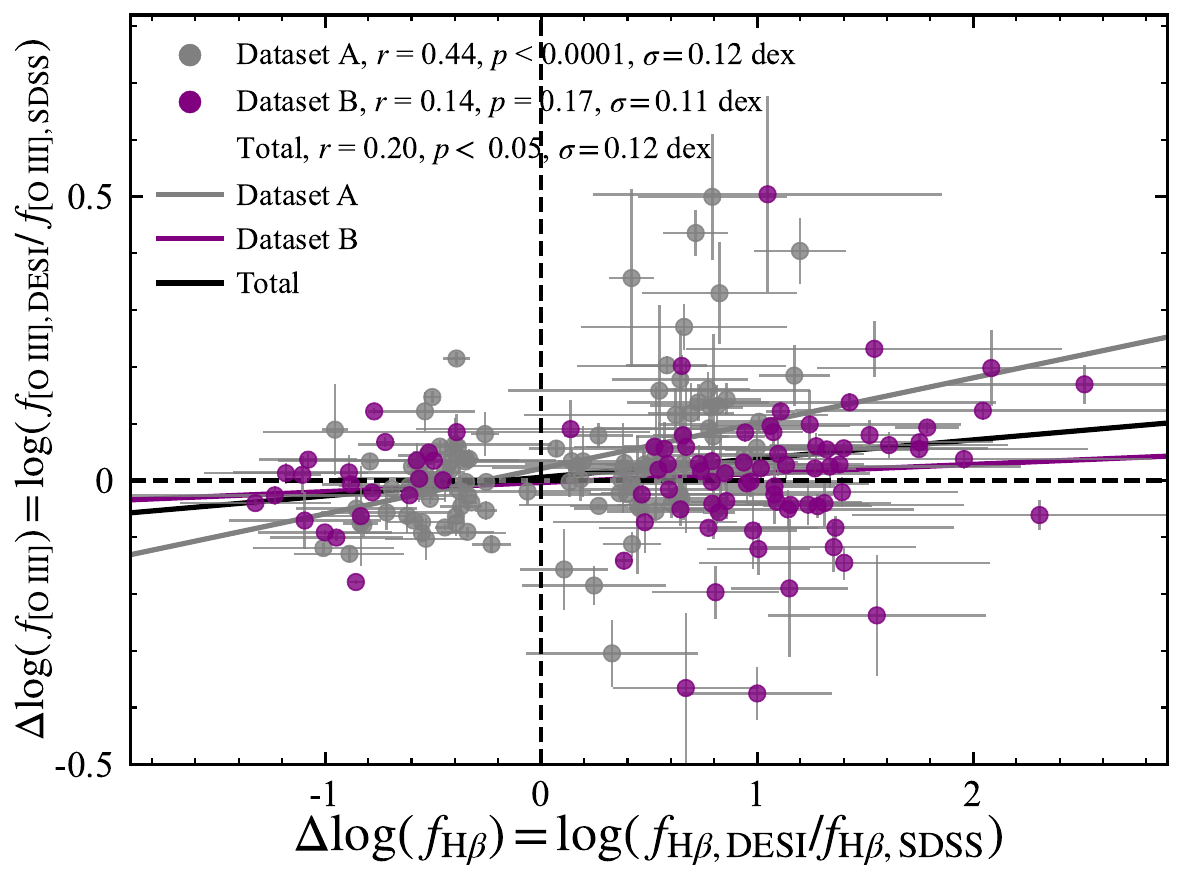}
\caption{
Relation between variations in [O~{\sc iii}]~$\lambda$5007 and broad H$\beta$ fluxes for the Total sample, Dataset A, and Dataset B. 
For Dataset A and the Total sample, [O~{\sc iii}]~$\lambda$5007 emission increases with increasing broad-line emission 
($r=0.44$, $p<0.0001$ for Dataset A; $r=0.20$, $p<0.05$ for Total sample), 
while this trend is not significant in Dataset B ($r=0.14$, $p=0.17>0.05$). 
Linear regression models illustrate these potential correlations. 
The scatter in [O~{\sc iii}]~$\lambda$5007 fluxes and the correlation coefficients (with $p$-values) 
are labeled in the figure. 
} 
\label{f5007}
\end{figure}

\begin{deluxetable*}{lcccccccccccc}
\tablewidth{0pt}
\setlength{\tabcolsep}{3pt} 
\tablecaption{Spectral parameters of 203 AGNs\label{tab:para}}
\tablehead{
\colhead{Epoch} & \colhead{RA} & \colhead{DEC} & \colhead{z} & \colhead{MJD}   & \colhead{$f_{\rm 5100}$} & \colhead{$\alpha$} & \colhead{$f_{\rm H\alpha}$} 
& \colhead{$f_{\rm H\beta}$} & \colhead{$f_{\rm [O~III]}$} & \colhead{$\rm FWHM_{H\alpha}$} & \colhead{$\rm FWHM_{\rm [O~III]}$} \\
\colhead{(1)} & \colhead{(2)} & \colhead{(3)} & \colhead{(4)} & \colhead{(5)}   & \colhead{(6)} & \colhead{(7)} & \colhead{(8)} & \colhead{(9)} & \colhead{(10)} & \colhead{(11)} & \colhead{(12)}
}
\startdata
SDSS & 00:13:45.67&+29:21:29.58& 0.094 & 56273.00 &$0.045 \pm 0.063$&$  -0.33\pm0.40$&$ 4.514  \pm3.451 $&$1.174  \pm1.317 $&$ 4.845 \pm0.106 $&  ---  &$ 365 \pm 12$ \\ 
DESI & 00:13:45.68&+29:21:29.59& 0.095 & 59503.15 &$0.069 \pm 0.072$&$  -1.27\pm0.95$&$ 8.064  \pm1.412 $&$3.436  \pm0.865 $&$ 4.578 \pm0.139 $&$ 3226  \pm 251   $&$ 319 \pm 2$  \\ \cline{1-12}
SDSS & 00:17:02.88&+25:05:59.65& 0.244 & 56243.00 &$0.032 \pm 0.003$&$  -1.57\pm0.31$&$ 7.219  \pm0.315 $&$2.480  \pm0.015 $&$ 0.697 \pm0.019 $&$ 6085  \pm 164   $&$ 311 \pm 15$ \\
DESI & 00:17:02.88&+25:05:59.68& 0.244 & 59523.21 &$0.009 \pm 0.006$&$  -0.87\pm0.68$&$ 0.678  \pm0.149 $&$0.207  \pm0.108 $&$ 0.758 \pm0.006 $& --- &$ 287 \pm 14$  \\ \cline{1-12}
SDSS & 00:23:26.10&+28:21:12.81& 0.243 & 57361.00 &$0.062 \pm 0.015$&$  -0.19\pm0.17$&$ 15.755 \pm0.973 $&$3.677  \pm0.171 $&$ 8.182 \pm0.040 $& --- &$ 462 \pm 1$ \\
DESI & 00:23:26.09&+28:21:12.89& 0.243 & 59503.18 &$0.297 \pm 0.032$&$  -1.51\pm0.33$&$ 77.041 \pm2.092 $&$23.947 \pm1.228 $&$ 7.290 \pm0.447 $&$ 9600  \pm 172   $&$ 409 \pm 27$  \\ \cline{1-12}
... & ... & ...  & ...  &  ... & ... & ...  &  ... &  ... &  ... &  ... &  ...  \\ \cline{1-12}
SDSS & 22:50:20.26&+01:32:17.87& 0.125 & 55500.00 &$0.005 \pm 0.006$&$  -0.47\pm0.94$&$ 8.860  \pm3.590 $&$0.206  \pm0.142 $&$ 1.553 \pm0.134 $&--- &$ 349 \pm 32 $ \\
DESI & 22:50:20.27&+01:32:17.94& 0.125 & 59481.25 &$0.035 \pm 0.041$&$  -0.17\pm0.35$&$ 11.407 \pm2.321 $&$0.766  \pm0.399 $&$ 1.765 \pm0.113 $&$ 4757  \pm 140   $&$ 324 \pm 48 $ \\ \cline{1-12}
SDSS & 23:10:56.05&+01:19:23.25& 0.197 & 55501.00 &$0.103 \pm 0.017$&$  -0.35\pm0.24$&$ 23.256 \pm2.159 $&$6.472  \pm0.292 $&$ 1.408 \pm0.021 $&$ 10446 \pm 375   $&$ 519 \pm 4$ \\
DESI & 23:10:56.05&+01:19:23.23& 0.197 & 59484.27 &$0.039 \pm 0.036$&$  -0.10\pm0.16$&$ 8.158  \pm1.068 $&$1.878  \pm1.000 $&$ 1.863 \pm0.143 $& --- &$ 503 \pm 19 $ 
\enddata
\tablecomments{
Col. (1) indicates the source of the epoch spectrum. Cols. (2–3) list the coordinates of object. Col. (4) gives the redshift. Col. (5) provides the Modified Julian Day. 
Col. (6) lists the optical continuum flux at 5100~\AA~in units of $\rm 10^{-15}~erg~s^{-1}~cm^{-2}~\AA^{-1}$. Col. (7) presents the spectral index. 
Cols. (8), (9), and (10) give the broad H$\alpha$, H$\beta$, and [O~{\sc iii}]~$\lambda$5007 fluxes in units of $\rm 10^{-15}~erg~s^{-1}~cm^{-2}$, respectively. 
Cols. (11) and (12) provide the full width at half maximum (FWHM) of the broad H$\alpha$ and [O~{\sc iii}]~$\lambda$5007 emission lines. 
The FWHM of broad H$\alpha$ is measured only from the high-luminosity epoch spectrum (i.e., with stronger H$\alpha$ flux) for each object; 
thus, the FWHM values for the low-luminosity epochs are marked as `---'. 
A horizontal line is used to separate one object from another in the table. \\
(This table is available in its entirety in machine-readable form in the online article.)
}
\end{deluxetable*}

\begin{deluxetable}{lcccc}
\tablewidth{0pt}
\tablecaption{Type transition statistics of 203 AGNs\label{tab:stat}}
\tablehead{
\colhead{Type Transition} & \colhead{Count ($\%$)} & \colhead{$\Delta$MJD [min, max]} & \colhead{Dataset} & \colhead{Notes}\\
\colhead{(1)} & \colhead{(2)} & \colhead{(3)} & \colhead{(4)} & \colhead{(5)}
}
\startdata
Type 1.0 $\leftrightarrow$ 1.0 & 4 (2.0\%) & [3662, 7697] &A &Not belonging to Turn-on/off\\
Type 1.0 $\leftrightarrow$ 1.2 & 6 (3.0\%) & [6205, 7275] &A&Not belonging to Turn-on/off\\
Type 1.0 $\leftrightarrow$ 1.5 & 24 (11.8\%) & [2560, 7900] &B&Belongs to Turn-on/off\\
Type 1.0 $\leftrightarrow$ 1.8/2.0 & 8 (3.9\%) & [2528, 7829] &B&Belongs to Turn-on/off\\
Type 1.2 $\leftrightarrow$ 1.2 & 4 (2.0\%) & [2194, 7219] &A&Not belonging to Turn-on/off\\
Type 1.2 $\leftrightarrow$ 1.5 & 65 (32.0\%) & [995, 7984] &A&Not belonging to Turn-on/off\\
Type 1.2 $\leftrightarrow$ 1.8/2.0 & 29 (14.3\%) & [1410, 7813] &B&Belongs to Turn-on/off\\
Type 1.5 $\leftrightarrow$ 1.5 & 27 (13.3\%) & [1548, 8075] &A&Not belonging to Turn-on/off\\
Type 1.5 $\leftrightarrow$ 1.8/2.0 & 32 (15.8\%) & [1817, 7667] &B&Belongs to Turn-on/off\\
Type 1.8/2.0 $\leftrightarrow$ 1.8/2.0 & 4 (2.0\%) & [5810, 7713] &A&Not belonging to Turn-on/off
\enddata
\tablecomments{
AGN Type transitions are categorized into distinct cases (Col.~1). 
Transition ratios among AGN subtypes and the shortest and longest observed intervals are shown in Col.~(2) and Col.~(3). 
Type transitions are classified into Dataset A, representing general AGNs with minor subtype changes, 
and Dataset B, representing significant transitions as typical turn-on/turn-off AGNs (Col.~4 and 5). 
}
\end{deluxetable}

\begin{figure*}[htb]
\centering
\includegraphics[angle=0,width=0.999\textwidth]{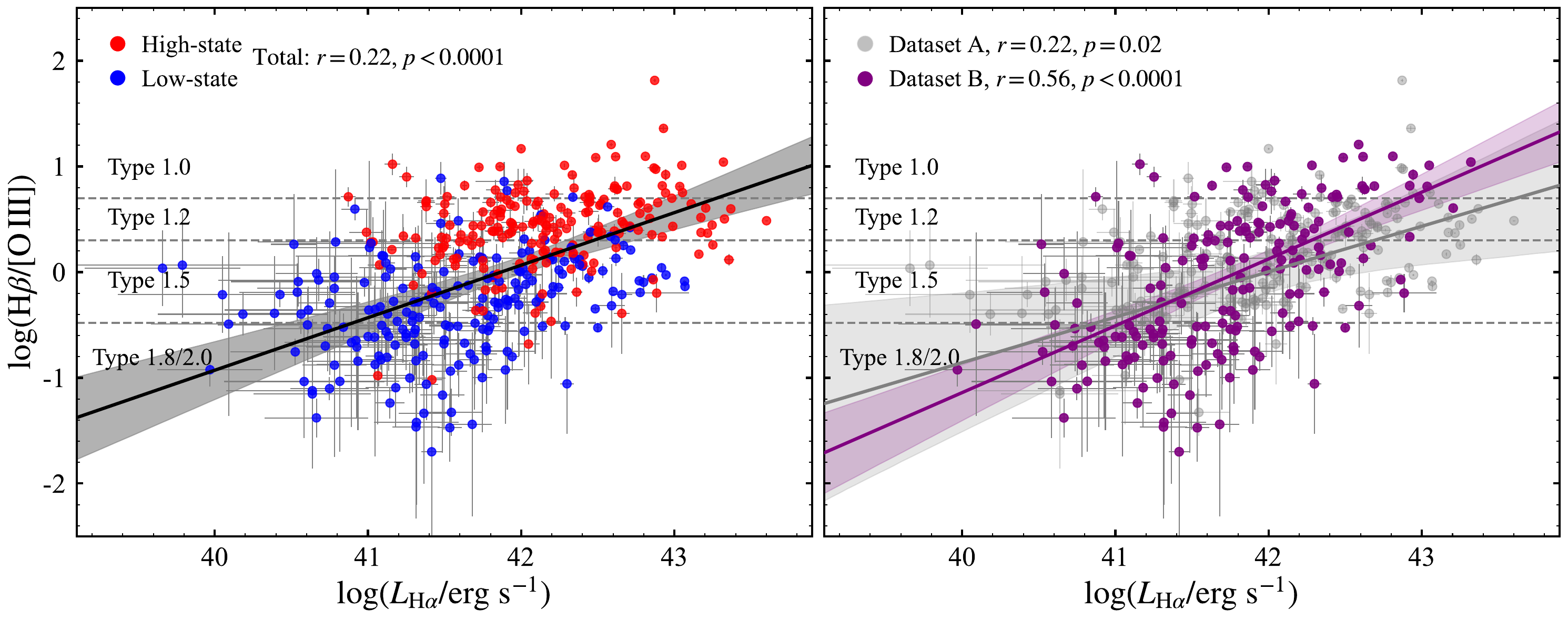}
\caption{
The relationship between the broad H$\beta$/[O~{\sc iii}]~$\lambda$5007 ratio (a proxy for AGN Types) 
and broad H$\alpha$ luminosity ($L_{\rm H\alpha}$) in logarithmic space for the Total sample (left panel) and Datasets A and B (right panel). 
Circles indicate measurements from SDSS and DESI epoch spectra. 
In the left panel, red circles show high-luminosity states, and blue circles show low-luminosity states. 
The black solid line, with its uncertainties, shows the linear regression result. 
In the right panel, gray and purple circles represent Datasets A and B, respectively. 
Using the broad H$\beta$ to [O~{\sc iii}]~$\lambda$5007 flux ratio and standard classification criteria (\citealt{Osterbrock1977,Winkler1992}), 
we label AGN Types 1.0, 1.2, 1.5, and 1.8/2.0 in regions separated by horizontal dashed lines. 
Regression results, including models and uncertainties, are shown for Dataset A (gray) and Dataset B (purple). 
The Pearson correlation coefficient $r$ and $p$-value are labeled.} 
\label{sedtype}
\end{figure*}

\begin{figure}[htb]
\centering
\includegraphics[angle=0,width=0.48\textwidth]{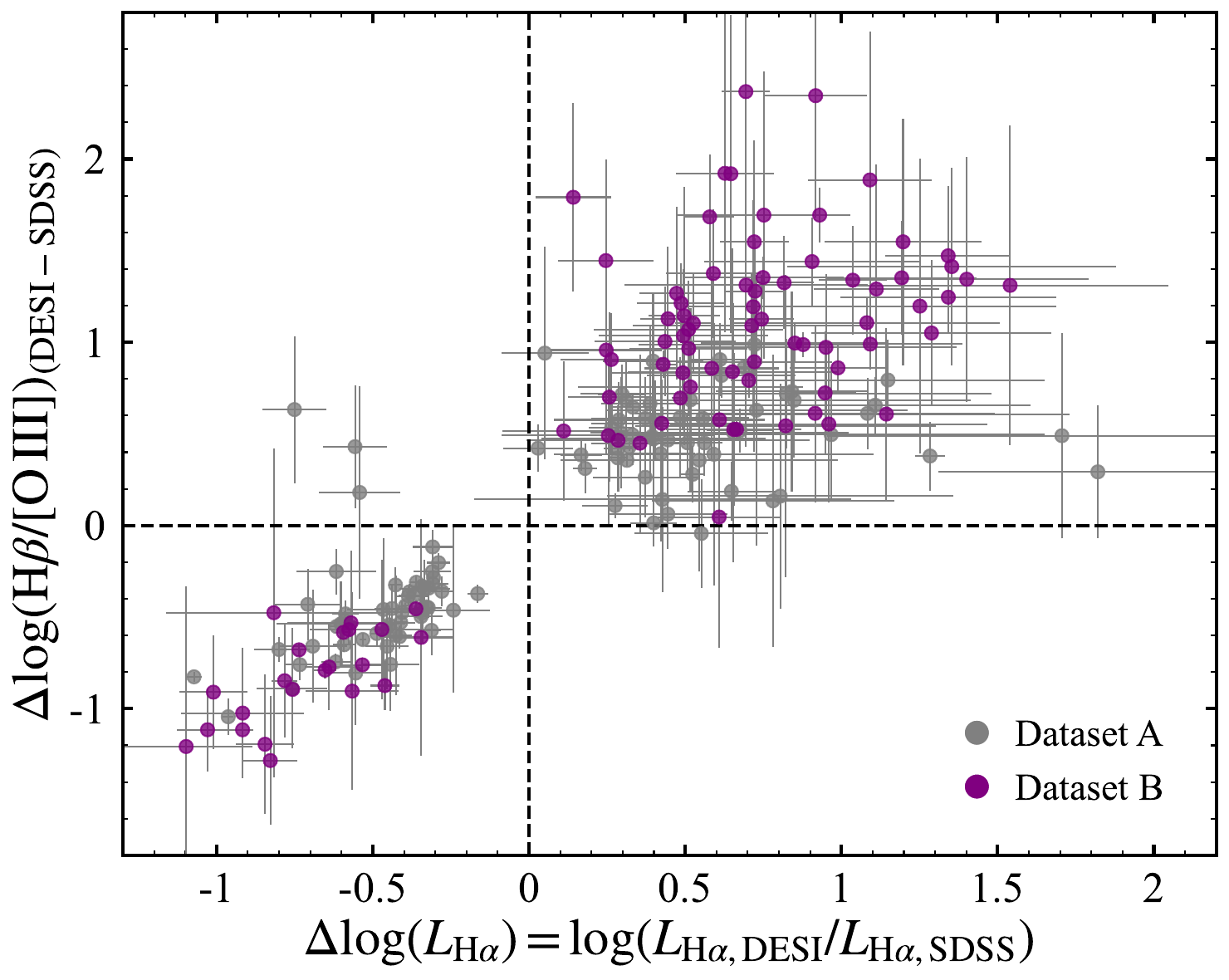}
\caption{
Relation between $\Delta$log(H$\beta$/[O~{\sc iii}]$~\lambda$5007) and $\Delta$log($L_{\rm H\alpha}$). 
Dataset A is shown in gray, Dataset B is presented in purple.} 
\label{dsedtype}
\end{figure}

\section{Parameter estimates and sample properties}\label{result1} 
\subsection{Spectral parameters}\label{pm} 
Our spectral fitting procedure shows that model uncertainty constitutes the primary source of systematic error 
in spectral decomposition measurements. This effect is particularly significant in low-luminosity AGN spectra, 
where the uncertainty of the host-galaxy template is the dominant factor. 
To assess its impact on spectral parameters, 
we retain fitting results obtained using multiple host-galaxy templates, 
specifically stellar templates with an age of 11 Gyr and metallicities of 0.008, 0.02, and 0.05 for each spectrum. 

For each spectrum, we measured the integrated fluxes of broad and narrow hydrogen Balmer lines, 
[O~{\sc iii}]~$\lambda$5007, and optical continuum at 5100~\AA, as well as the full width at half maximum (FWHM) 
of the broad H$\alpha$ and [O~{\sc iii}]~$\lambda$5007 line, from the best-fitted models.  
The average value and standard deviation of each parameter were adopted as the best estimates and corresponding systematic uncertainties. 
The Poisson errors for flux measurements were determined directly from the spectral errors. 
All derived spectral parameters, along with basic object information, are summarized in Table~\ref{tab:para}. 
To assess the reliability of our measurements, 
we estimated the measurement accuracy for three key quantities: the optical continuum flux, 
and the broad H$\beta$ and H$\alpha$ line fluxes for each epoch spectrum. 
The resulting median measurement accuracies are 44\%, 19\%, and 11\%, respectively. 
The relatively low precision of the optical continuum flux (44\%) is consistent with expected contamination from host-galaxy starlight, 
supporting our earlier finding that host-galaxy template uncertainty is the dominant systematic factor.

Notably, some CL-AGN spectra exhibit a clear broad H$\alpha$ signal but lack a detectable broad H$\beta$ component, 
making it impossible to include the broad H$\beta$ component in the spectral fitting (i.e., it was fitted as zero). 
In these cases, we estimated the broad H$\beta$ flux from the fitting residuals.  

\subsection{Sample properties}\label{sprop} 
According to the ratio of broad H$\beta$ to [O~{\sc iii}]~$\lambda$5007 line flux and the classification criteria outlined in Section~\ref{intro}, 
we find that 11.3\% of the spectra are classified as Type 1.0, 26.6\% as Type 1.2, 43.1\% as Type 1.5, and 19\% as Type 1.8/2.0 in this sample. 
Then we categorize AGN type transitions into 10 distinct cases and report their respective proportions in the sample, 
along with the shortest and longest observed time intervals, as shown in Table~\ref{tab:stat}. 
Furthermore, we divide the Total sample into two datasets: Dataset A, characterized by minor type variations (110 AGNs), 
potentially representing general AGN variability; 
and Dataset B, exhibiting significant type transitions (93 AGNs), 
which display the typical turn-on/turn-off features of CL-AGNs. 

In Figure~\ref{pl}, we examine variations in the spectral index ($\alpha$), the optical continuum flux at 5100~\AA\ ($f_{\rm 5100}$), 
and the broad H$\alpha$ flux ($f_{\rm H\alpha}$) for the Total sample, and both datasets, using previously measured parameters. 
The standard deviations are 0.56 dex for $\Delta \alpha$=$\alpha_{_{\rm DESI}}-\alpha_{_{\rm SDSS}}$, 
0.33 dex for $\Delta$log($f_{\rm 5100}$)=log($f_{\rm 5100,DESI}/f_{\rm 5100,SDSS}$), 
and 0.66 dex for $\Delta$log($f_{\rm H\alpha}$)=log($f_{\rm H\alpha,DESI}/f_{\rm H\alpha,SDSS}$), 
indicating significant variability in the spectral shape, AGN continuum emission, and the broad line fluxes. 
We analyze the relationships between $\Delta \alpha$ and $\Delta$log($f_{\rm 5100}$), 
and between $\Delta \alpha$ and $\Delta$log($f_{\rm H\alpha}$), 
and find that both “bluer-when-brighter” and “redder-when-brighter” trends are present in the sample. 

Additionally, as shown in Figure~\ref{pl}, many variations in $f_{\rm 5100}$ are close to 0 (indicating no variability), 
whereas variations in $f_{\rm H\alpha}$ clearly deviate from 0. When AGNs exhibit clear type transitions, 
particularly in Dataset B with significant type transition, strong emission line variability is expected as seen in the right panel. 
Under such conditions, we would expect $f_{\rm 5100}$ to vary strongly, with amplitude comparable to or exceeding that of $f_{\rm H\alpha}$. 
This expectation is supported by numerous reverberation mapping studies (e.g., \citealt{Du2015,Lu2021,Lu2022,Woo2024,Hu2025}), 
which show that $f_{\rm 5100}$ typically varies as strongly as or more strongly than Balmer broad lines. 
This contrast reinforces the findings/limitations in Figure~\ref{fig:specfit} (also see Section~\ref{specfit}), 
where the strong degeneracy between the AGN power-law continuum and host-galaxy starlight 
leads to large measurement uncertainties in $f_{\rm 5100}$. 
Combining the measurement accuracy of the optical continuum flux is significantly lower than that of the broad emission line, 
we prioritize H$\alpha$ luminosity over optical continuum luminosity in subsequent analyses when viable alternatives exist. 

We investigate the relation between variations in [O~{\sc iii}]~$\lambda$5007 and broad H$\beta$ fluxes across our sample. 
The [O~{\sc iii}] flux variation is quantified as $\Delta$log($f_{\rm [O~III]}$)=log($f_{\rm [O~III],DESI}/f_{\rm [O~III],SDSS}$), 
and broad H$\beta$ variation is defined as $\Delta$log($f_{\rm H\beta}$)=log($f_{\rm H\beta,DESI}/f_{\rm H\beta,SDSS}$). 
These relationships were examined separately for the Total sample, Dataset A, and Dataset B, as presented in Figure~\ref{f5007}. 
Correlation analysis reveals distinct behaviors across the different datasets: 
Dataset A shows a statistically significant positive correlation 
(Pearson coefficient: $r=0.44$, null hypothesis probability: $p<0.0001$) with low dispersion of 0.12~dex. 
Dataset B shows no significant correlation ($r=0.14$, $p=0.17>0.05$), despite demonstrating nearly identical dispersion (0.11~dex) to Dataset A. 
Total sample displays a weak overall correlation ($r=0.20$, $p<0.05$), 
which likely reflects an averaging effect between the two distinct dataset trends. 
We simply perform a linear regression using the Python package `linmix'
\footnote{\url{https://linmix.readthedocs.io/en/latest/src/linmix.html}} 
between $\Delta$log($f_{\rm [O~III]}$) 
and $\Delta$log($f_{\rm H\beta}$) to illustrate the potential correlations, as shown in Figure~\ref{f5007}. 
More comprehensive regression analyses will be presented in Section~\ref{ra}. 
The observed correlation may arise because, over timescales of several thousand days (see Table~\ref{tab:stat}), 
the kpc-scale narrow-line region (NLR) has sufficient time to respond to nuclear luminosity changes during AGN type transitions. 
All $p$-values reported here and henceforth follow statistical reporting guidelines \citep{Lazzeroni2014}; 
Extremely small $p$ values are presented in the form of an upper limit (p$<$0.0001) rather than the exact value to avoid conveying false precision. 

\subsection{Broad H$\alpha$ luminosity and Eddington ratio}\label{edd} 
Estimating the Eddington ratio from a single-epoch spectrum requires reliable measurements 
of both the supermassive black hole (SMBH) mass ($M_\bullet$) and bolometric luminosity ($L_{\rm bol}$). 
The virial method is commonly used to estimate $M_\bullet$ from single epoch spectrum:  
$M_\bullet = f R_{\rm BLR} V^{2}/G$, 
where $R_{\rm BLR}$ denotes the radius of the broad-line region (BLR), 
$V$ is the line width (e.g., FWHM of the broad emission line), 
$G$ is the gravitational constant, and $f$ is a dimensionless factor accounting for the geometry and kinematics of BLR. 

The optical monochromatic luminosity at 5100~\AA\ ($L_{\rm 5100}$) is usually used to 
estimate the BLR radius of $R_{\rm BLR}$ via the H$\beta$-based BLR size-luminosity relation (\citealt{Bentz2013}) 
and to derive the accretion rate or Eddington ratio based on accretion disk models (\citealt{Du2014,Du2018b}). 
However, due to the blending of host-galaxy starlight and AGN continuum, particularly in low-luminosity AGNs, 
the measurements of $f_{\rm 5100}$ estimates are subject to large uncertainties, 
which propagate into estimates of the BLR radius and Eddington ratio. 
The median measurement accuracy of $f_{\rm 5100}$ for the total sample is 44\% (see Section~\ref{pm}), 
more considerations or explanations please refer to Section~\ref{sprop}. 
In contrast, the broad H$\alpha$ luminosity ($L_{\rm H\alpha}$) can be measured 
more reliably through spectral fitting and decomposition 
(median measurement accuracy: 11\% for the total sample, see Section~\ref{pm}), 
as it is less contaminated by host-galaxy starlight (\citealt{Stern2012a,Stern2012b,Ho2008}). 
Moreover, in many CL-AGNs, especially in low-luminosity states, 
the broad H$\beta$ emission is weak or undetectable, making FWHM measurements difficult or unreliable. 

Therefore, we adopt the H$\alpha$-based BLR size-luminosity relation (see Equation 3 of \citealt{Cho2023}, \citealt{Wu2004}) 
to estimate the BLR radius, and use the broad H$\alpha$ FWHM as the BLR velocity. 
Assuming that the intrinsic width of the [O~{\sc iii}]~$\lambda$5007 line remains unchanged 
between the SDSS and DESI epochs, we correct for instrumental resolution differences 
by applying the observed changes in [O~{\sc iii}]~$\lambda$5007 line width to the broad H$\alpha$ FWHM measurements. 

We then combine the empirical relation between $L_{\rm 5100}$ and $L_{\rm H\alpha}$ (\citealt{Greene2005}):  
$L_{\rm 5100}=2.39\times10^{43}(L_{\rm H\alpha}/10^{42})^{0.86}{\rm erg~s^{-1}}$, 
with the bolometric correction $L_{\rm bol} = 9.8 \times L_{\rm 5100}$ (\citealt{McLure2004}), 
to derive the bolometric luminosity as a function of $L_{\rm H\alpha}$: 
$L_{\rm bol} = 2.34\times10^{44}(L_{\rm H\alpha}/10^{42})^{0.86}~\rm erg~s^{-1}$. 
Using these relations, we compute the SMBH mass for each CL-AGN 
based on the spectrum (SDSS or DESI) with the strongest broad H$\alpha$ flux (i.e., during the high-luminosity state), 
and estimate Eddington ratio of $\eta=L_{\rm bol}/L_{\rm Edd}$ for each epoch, 
where the Eddington luminosity $L_{\rm Edd}$=$1.26\times10^{38}(M_\bullet/M_{\sun})$~erg~s$^{-1}$. 

\section{Regression analysis} \label{ra}
\subsection{AGN Type and broad H$\alpha$ luminosity} \label{ra1} 
\cite{Stern2012a} reported a strong linear relationship between the optical-ultraviolet spectral energy distribution (SED) 
and broad H$\alpha$ luminosity ($L_{\rm H\alpha}$), suggesting that variations in $L_{\rm H\alpha}$ reflects changes in the SED. 
This implies that fluctuations in $L_{\rm H\alpha}$ are indicative of changes in the ionizing flux within the broad-line region.

Motivated by this finding, we examine the relationship between the 
broad H$\beta$/[O~{\sc iii}] flux ratio (a proxy for AGN type) 
and $L_{\rm H\alpha}$ in logarithmic space, as shown in Figure~\ref{sedtype}, 
for the Total sample, Dataset A, and Dataset B. 

The left panel of Figure~\ref{sedtype} presents the results for the Total sample, 
red circles denote observations associated with high-luminosity states, 
while blue circles represent those from low-luminosity states, visually distinguishing the two populations. 
The data reveal a statistically significant positive correlation between H$\beta$/[O~{\sc iii}] and $L_{\rm H\alpha}$, 
with a Pearson coefficient of $r=0.22$ and a null hypothesis probability of $p<0.0001$, 
indicating that the observed correlation is not due to random chance. 

The right panel of Figure~\ref{sedtype} displays the analysis results of Dataset A and Dataset B, 
with gray and purple circles representing Dataset A and B, respectively. 
Statistical analysis show that Dataset A exhibits a marginal positive correlation 
between H$\beta$/[O~{\sc iii}] and $L_{\rm H\alpha}$ ($r=0.22$, $p=0.02<0.05$). 
In contrast, Dataset B dataset reveals a statistically robust positive correlation between 
H$\beta$/[O~{\sc iii}] and $L_{\rm H\alpha}$ ($r=0.56$, $p<0.0001$). 

To quantify these correlations, we performed a linear regression analysis between 
log(H$\beta$/[O~{\sc iii}]) and log($L_{\rm H\alpha}$) for the Total sample, Dataset A, and Dataset B. 
To account for measurement uncertainties in both variables, we employed the Python package `linmix', 
which implements the Bayesian linear regression framework proposed by \cite{Kelly2007}. 
This method rigorously models uncertainties in both variables and estimates the regression parameters. 
The resulting best-fit models yield the slope, intercept, intrinsic scatter, and associated uncertainties. 
which are plotted in Figure~\ref{sedtype} and summarize below: 

\begin{equation}
\begin{split}
&{\rm log(\frac{H\beta}{[O~III]})} =  \\
&\left\{
\begin{aligned}
(0.50\pm 0.07) {\rm log}(L_{\rm H\alpha})-(20.74\pm2.86)  \pm 0.79 \\\text{(for  Total sample),} \\
(0.43\pm 0.16) {\rm log}(L_{\rm H\alpha})-(18.19\pm6.66)  \pm 1.45 \\\text{(for Dataset A),} \\
(0.63\pm 0.07) {\rm log}(L_{\rm H\alpha})-(26.49\pm2.96)  \pm 0.48 \\\text{(for Dataset B).} 
\end{aligned}
\right.
\end{split}
\label{Eq1}
\end{equation}

Next, we explore additional observational features. In the left panel of Figure~\ref{sedtype}, 
we observe that high-luminosity-state AGNs exhibit a flattening trend in the relation between 
log(H$\beta$/[O~{\sc iii}]) and log($L_{\rm H\alpha}$). 
This behavior may arise from variations in either [O~{\sc iii}]~$\lambda$5007 or broad H$\beta$.  

Regarding broad H$\beta$, multiple studies (\citealt{Bon2018, Gaskell2021, Panda2022, Homan2023}) 
have shown that H$\beta$ luminosity increases with continuum luminosity, 
following the “Pronik-Chuvaev effect” first identified by \cite{Pronik1972}. 
This effect describes a flattening of the H$\beta$ response at high luminosities, 
where the H$\beta$ emission tends to saturate (see Figure 3 of \citealt{Panda2022}, Figure 4 of \citealt{Panda2023}). 
\cite{Panda2022} further noted that this phenomenon is particularly evident in 
Population B AGNs with low-luminosity states ($L_{\rm bol}/L_{\rm Edd} \leq 0.2$). 
If present in our sample, such saturation could flatten the 
log(H$\beta$/[O~{\sc iii}])–log($L_{\rm H\alpha}$) relation 
at high $L_{\rm H\alpha}$ or weaken the overall correlation. 

For [O~{\sc iii}]~$\lambda$5007, we found in Section~\ref{sprop} that its flux 
slightly increases with enhanced nuclear activity over time intervals of 1000$\sim$8000 days across the Total sample. 
This trend is prominent in Dataset A and likely contributes to the weaker correlation observed there. 
In contrast, it is less pronounced in Dataset B, 
where the stronger correlation between log(H$\beta$/[O~{\sc iii}]) and log($L_{\rm H\alpha}$) persists 
(see the correlation and regression analyses presented above). 
Collectively, our results do not show evidence of H$\beta$ saturation in the current sample. 
At minimum, the correlation and regression analysis results for both Dataset A and Dataset B support this interpretation. 

However, the above correlation analysis represents an ensemble average across the sample. 
Due to overlap between high- and low-state measurements, 
individual source variability patterns are obscured. 
Moreover, combining multi-epoch observations with varying time intervals dilutes intrinsic source-specific variability.

To address this, we analyze the relationship between $\Delta$log(H$\beta$/[O~{\sc iii}]) 
and $\Delta$log($L_{\rm H\alpha}$) in Figure~\ref{dsedtype}, 
aiming to uncover how these variations are connected to source variability. 
We find that, except for four points (from Dataset A), 
all data pairs [$\Delta$log(H$\beta$/[O~{\sc iii}]), $\Delta$log($L_{\rm H\alpha}$)] 
lie in the first or third quadrants. This distribution suggests a consistent variation trend: 
increases in H$\beta$/[O~{\sc iii}] accompany increases in $L_{\rm H\alpha}$ (first quadrant), 
and decreases correspond to decrease (third quadrant). 

Given the established link between $L_{\rm H\alpha}$ and optical-ultraviolet SED (\citealt{Stern2012a}), 
these results support the hypothesis that AGN type transitions are initially driven by SED variations. 
Furthermore, they provide direct observational evidence for the theoretical prediction that 
transitions among AGN Types (1.0 to 2.0) are influenced by changes 
in the ionizing photon flux within the BLR (\citealt{Korista2004}). 

\begin{figure*}[htb]
\centering
\includegraphics[angle=0,width=0.999\textwidth]{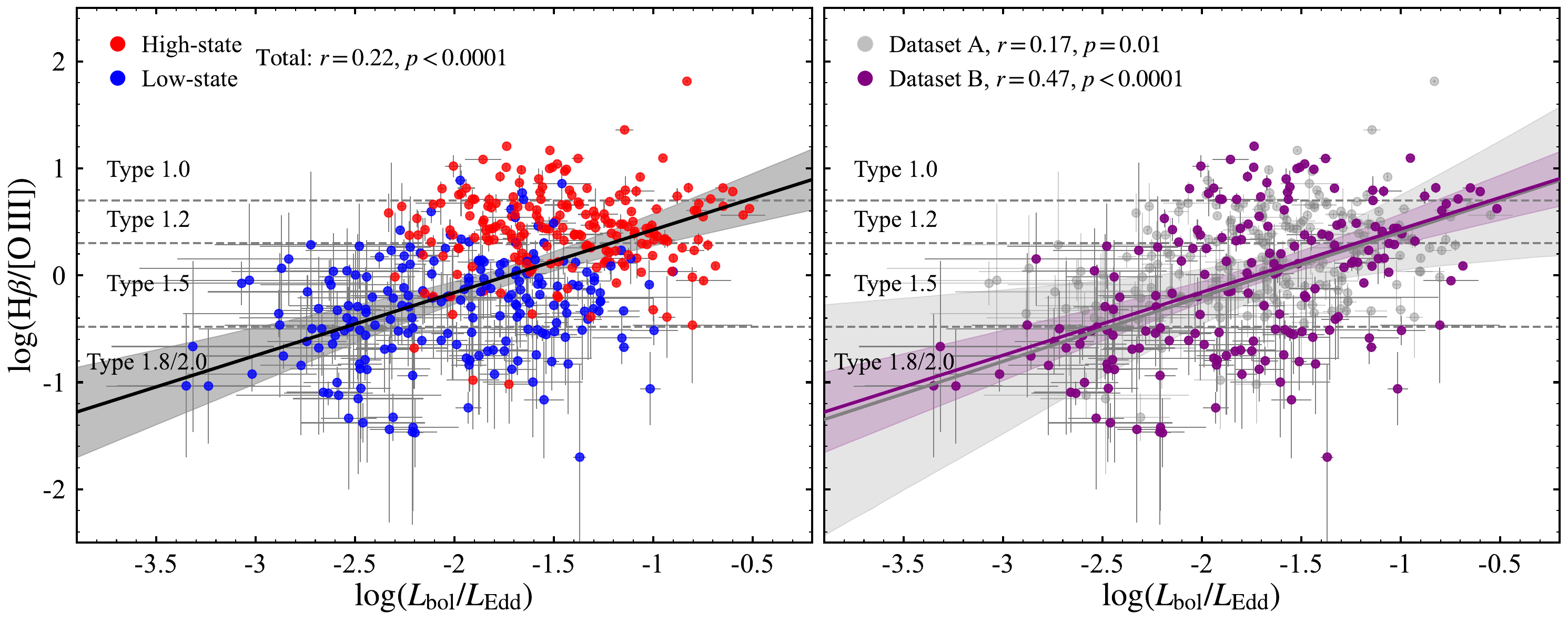}
\caption{
Same as Figure~\ref{sedtype}, but for the relationship between H$\beta$/[O~{\sc iii}]~$\lambda$5007 
and Eddington ratio ($L_{\rm bol}/L_{\rm Edd}$). 
\label{eddtype}}
\end{figure*}

\begin{figure}[htb]
\centering
\includegraphics[angle=0,width=0.48\textwidth]{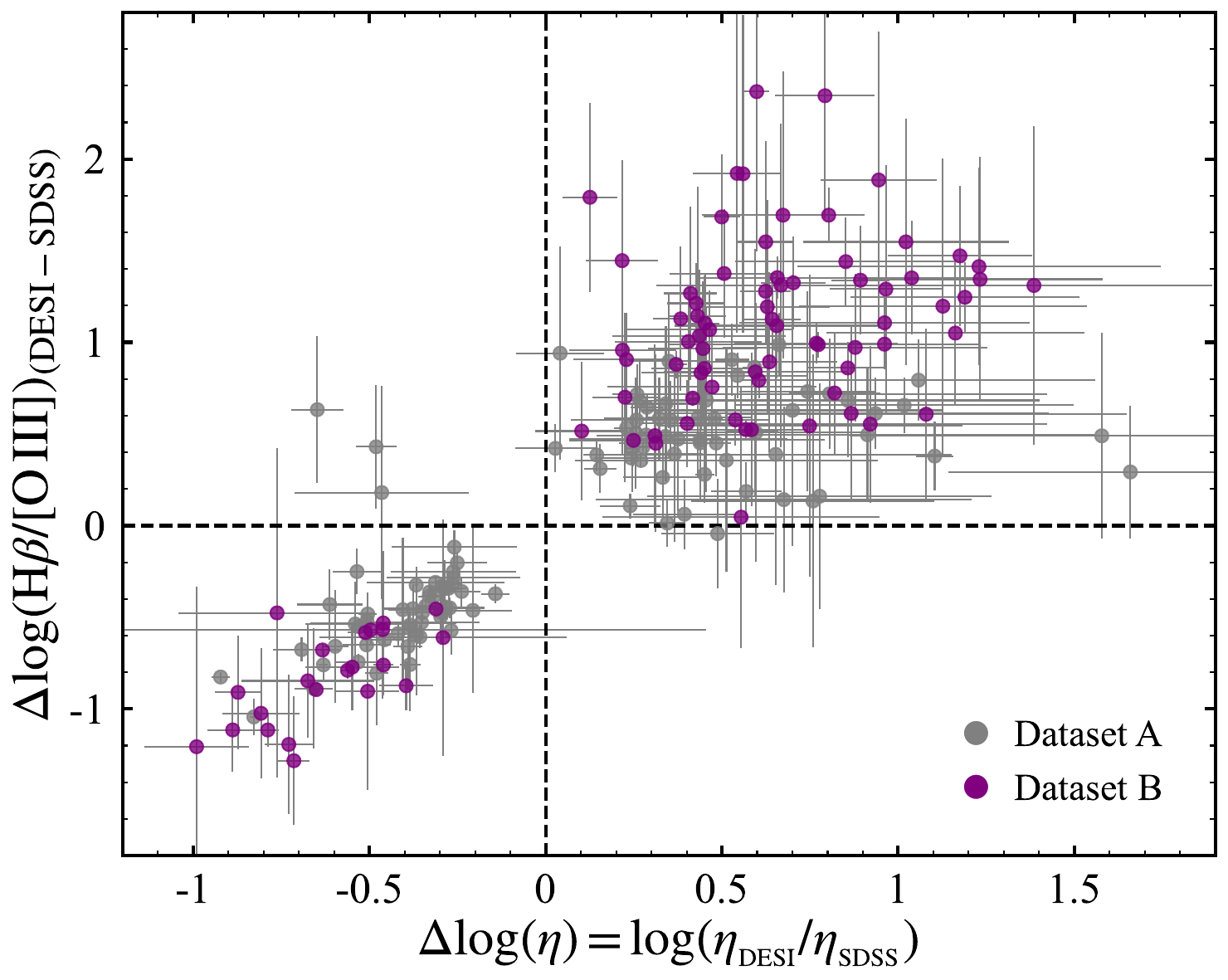}
\caption{
Same as Figure~\ref{dsedtype}, but for the relation between $\Delta$log(H$\beta$/[O~{\sc iii}]~$\lambda$5007) and $\Delta$log($\eta$), 
where $\eta=L_{\rm bol}/L_{\rm Edd}$. 
} 
\label{deddtype}
\end{figure}

\subsection{AGN Type and Eddington ratio} \label{ra2} 
Motivated by the recently observed correlation between the type transition of Mrk~1018 and its Eddington ratio 
(\citealt{Lu2025}), along with similar findings reported in \cite{Panda2024} and \cite{Jana2025}, 
we conduct a further investigate into the underlying mechanism driving AGN type transitions using our selected CL-AGN sample. 
Similar to the analysis in Section~\ref{ra1}, we examine the relationship between H$\beta$/[O~{\sc iii}] 
and $L_{\rm bol}/L_{\rm Edd}$ in logarithmic space for the Total sample, Dataset A, and Dataset B, as shown in Figure~\ref{eddtype}. 
The left panel of Figure~\ref{eddtype} displays the results for the Total sample, 
where red circles represent high-luminosity states and blue circles denote low-luminosity states. 
Correlation analysis reveals a statistically significant positive correlation 
between H$\beta$/[O~{\sc iii}] and $L_{\rm bol}/L_{\rm Edd}$ 
with a Pearson coefficient of $r = 0.44$ a null hypothesis probability of $p<0.0001$. 

We also analyze this relationship for Dataset A and Dataset B (right panel of Figure~\ref{eddtype}), 
using the same symbol scheme as in Figure~\ref{sedtype}. 
Dataset A shows a marginal positive correlation ($r=0.17$, $p=0.01<0.05$), 
while Dataset B exhibits a robust positive correlation ($r=0.47$, $p<0.0001$). 

To better characterize these correlations, 
we perform linear regression analysis between 
log(H$\beta$/[O III]) and log($L_{\rm bol}/L_{\rm Edd}$) 
using the same Bayesian method described in Section~\ref{ra1}. 
The best-fit models are shown in Figure~\ref{eddtype} and summarized as: 

\begin{equation}
\begin{split}
&{\rm log(\frac{H\beta}{[O~III]})} =  \\
&\left\{
\begin{aligned}
(0.59\pm 0.09) {\rm log}(L_{\rm bol}/L_{\rm Edd}) + (1.01\pm0.16) \pm 0.83 \\\text{(for Total sample),} \\
(0.60\pm 0.24) {\rm log}(L_{\rm bol}/L_{\rm Edd}) + (1.00\pm0.40) \pm 1.45 \\\text{(for Dataset A),} \\
(0.59\pm 0.08) {\rm log}(L_{\rm bol}/L_{\rm Edd}) + (1.02\pm0.15) \pm 0.53 \\\text{(for Dataset B).} \\
\end{aligned}
\right.
\end{split}
\label{Eq2}
\end{equation}

As in Figure~\ref{dsedtype} (also see Section~\ref{ra1}), 
we examine the relation between $\Delta$log(H$\beta$/[O~{\sc iii}])
and $\Delta$log($L_{\rm bol}/L_{\rm Edd}$), shown in Figure~\ref{deddtype}. 
Except for four points from Dataset A, all data pairs 
[$\Delta$log(H$\beta$/[O~{\sc iii}]), $\Delta$log($L_{\rm bol}/L_{\rm Edd}$)] 
lie in the first or third quadrants, indicating a consistent co-variation trend across sources: 
increases in one quantity accompany increases in the other, and decreases occur together. 

The regression results, specifically Equations~(\ref{Eq1}) and (\ref{Eq2}), 
clearly demonstrate that variations in the accretion rate regulate the ionizing flux within the BLR, 
thereby triggering AGN type transitions. 
Recent studies have provided indirect but converging evidence from multiple perspectives (e.g., \citealt{Noda2018,Sniegowska2020,Zeltyn2024,Panda2024,Jana2025}). 
A brief summary follows; for more details, see the review by \cite{Ricci2023}. 

\cite{Panda2024} analyzed multi-epoch spectra of 93 CL-AGN from SDSS/BOSS/eBOSS, 
focusing on their evolutionary paths along the quasar main sequence (Eigenvector 1, EV1), 
defined by the H$\beta$ broad-line FWHM and optical Fe~{\sc ii} strength (e.g., \citealt{Marziani2018,Panda2024FASS}). 
CL-AGNs are predominantly found at low Eddington ratios ($\sim$0.01 to 0.1), 
supporting the idea that accretion flow instability, 
such as transitions between standard disk and advection-dominated accretion flow (ADAF) phases, 
drives the CL phenomenon. Where five sources were observed to shift between Population A and B (\citealt{Sulentic2000}), 
demonstrating that abrupt changes in accretion can induce spectral type transitions. 
\cite{Zeltyn2024} analyzed 116 CL-AGNs and found that they typically occur 
at low Eddington ratios ($L_{\rm bol}/L_{\rm Edd} \approx 0.025$), 
suggesting their behavior is governed by accretion rate variability rather than orientation effects alone. 

\cite{Sniegowska2020} simulated the time-dependent instabilities evolution 
in the transition zone between the standard disk and ADAF, 
offering a self-consistent physical explanation for the recurrent outbursts observed in CL-AGN. 
Additionally, \cite{Noda2018} modeled the broad-band (optical/UV and X-ray) spectrum 
of Mrk 1018 during its transition from a Seyfert 1 to a Seyfert 1.9 over approximately 8 years. 
They found that the changing-look behavior arises from an inner disk state transition (thin disk to ADAF) 
coupled with a significant drop in outer disk accretion rate. 
Radiation pressure in AGN disks shortens the variability timescale, 
enabling significant changes within years. This model explains the majority 
of CL-AGNs and predicts their occurrence at $L_{\rm bol}/L_{\rm Edd} \approx 0.02\text{--}0.03$, consistent with observations. 

These results collectively show that AGN type transitions are fundamentally linked 
to changes in the accretion rate onto the central black hole. 
They support an evolving view of AGN unification that incorporates both viewing angle and accretion history.

Moreover, \cite{Giustini2019} proposed a framework for AGN inner regions 
based on black hole mass accretion rates and black hole masses, 
identifying five regimes with distinct accretion and ejection behaviors. 
These regimes correspond to different observed AGN types, 
from inactive galaxies to super-Eddington sources, with disk wind strength dependent on both parameters (also see \citealt{Nicastro2000}). 
The model elucidates how radiative and kinetic feedback vary across regimes and offers a foundation for interpreting observations. 
However, more quantitative research is needed to better define the boundaries between these regimes. 
Collectively, these findings reinforce the role of accretion evolution in driving AGN type transitions 
and support a dynamic unification model that integrates both accretion history and viewing angle. 

\section{Discussion}\label{dis}
In our previous analysis, we identified correlations between type transitions in CL-AGNs 
and variations in broad H$\alpha$ luminosity and the Eddington ratio. 
However, these two parameters only reflect the average effects of changes in the AGN spectral energy distribution (SED; \citealt{Stern2012a,Cai2023}). 
The detailed evolution of the SED during the transition phase in individual CL-AGNs, 
and how such object-specific SED variations influence the observed correlations, remain unclear. 
This question lies beyond the scope of the present study; addressing it will require multi-wavelength, 
quasi-simultaneous observations to construct time-resolved SEDs across a sample of sources. 
We anticipate that future investigations, 
through population-level studies or detailed case analyses, 
will explore this issue more deeply. 

To date, only a few CL-AGNs have shown indirect evidence of unusual SED evolution. 
For example, Mrk 1018 (\citealt{Lyu2021, Saha2025}) exhibits a `V-shaped' relation between the X-ray spectral index 
(or X-ray-uv-optical slope) and either the Eddington ratio or AGN luminosity. 
This behavior may indidate structural changes in the accretion disk near an Eddington ratio of $\sim$0.02 
(see \citealt{Ruan2019,Lyu2021,Saha2025}). Notably, this `V-shaped' trend corresponds to a color change 
in the accretion disk radiation, 
specifically, the source becomes brighter and bluer,  or brighter and redder. 

Although our current sample includes only two spectral epochs and thus cannot fully characterize the `V-shaped' relation, 
we examined the relationships between $\Delta \alpha$ and $\Delta$log($f_{\rm 5100}$), 
as well as between $\Delta \alpha$ and $\Delta$log($f_{\rm H\alpha}$). 
Despite the relatively low measurement precision for these parameter, 
we find that both “bluer-when-brighter” and “redder-when-brighter” trends coexist within the sample. 
In future work, we plan to obtain quasi-simultaneous, multi-wavelength data for additional objects to construct complete SED sequences, 
enabling a more comprehensive investigation into how detailed SED evolution influences the correlations reported in this study. 

On the other hand, the empirical relations derived in Section~\ref{ra} (Equation~\ref{Eq1} and~\ref{Eq2}) 
exhibit substantial intrinsic scatter, likely arising from multiple physical and observational factors: 
(1) The broad H$\beta$ line flux measurements have a relatively low accuracy ($\sim$20\%), 
potentially introducing dispersion into the broad H$\beta$/[O~{\sc iii}]~$\lambda$5007 ratio; 
(2) The estimation of the Eddington ratio in Equation~(\ref{Eq2}) relies on multiple empirical relations, 
each with its own intrinsic scatter. Combined with our measurement uncertainties, 
this can propagate additional dispersion into the derived relations; 
(3) Theoretically, recombination lines such as H$\beta$ exhibit nonlinear responses to continuum variations due to optical depth effects (\citealt{Korista2004}). Variations in line emissivity and responsivity with gas density may further contribute to the observed scatter; 
(4) The current sample cannot unambiguously identify the physical drivers of changes in accretion rate (Eddington ratio) or variability, 
such as tidal disruption events (TDEs, \citealt{Li2022,Zhang2022,Jiang2025}), obscuration by dusty clouds (\citealt{Ricci2023}), 
structural changes in the accretion disk (\citealt{Noda2018, Sniegowska2020, Saha2025}), or other mechanisms. 
The potential coexistence of multiple processes in the sample likely introduces additional dispersion; 
(5) Furthermore, star formation and stellar evolution in the self-gravitating regions of AGN disks 
have been widely proposed (e.g., \citealt{Cheng1999, Wang2021, Cantiello2021, Wang2023}). 
Energy injection from stellar processes or outbursts in the outer disk may represent 
an alternative physical mechanism contributing to the observed scatter. 
The potential influences of these mechanisms are need to be investigated in future studies. 

Overall, these findings suggest a prospect that increased ionizing fluxes within the BLR, 
driven by higher accretion rates, may play a key role in triggering AGN type transitions. 
Many studies have reached similar findings from diverse research perspectives. 
To enhance readers' understanding of these convergent findings, 
we have briefly summarized and provided examples of these studies in Section~\ref{ra2} 
(also see \citealt{Panda2024,Zeltyn2024,Sniegowska2020,Noda2018}). 

Additionally, \cite{Lu2025} found that the famous CL-AGN of Mrk~1018 undergoes 
full-cycle changing-look behavior accompanied by full-cycle type transitions. 
They further found strong evidence that the full-cycle type transition is regulated by accretion: 
Mrk~1018 evolved from Type 2.0 to 1.0 as its Eddington ratio increases dramatically, 
supporting the accretion-driven model for CL-AGNs (also see \citealt{Sheng2017,Lyu2021,Ricci2023,Saha2025}). 
The results in this study (Equation~\ref{Eq1} and \ref{Eq2}) 
offer a preliminary assessment of the generalizability of the Mrk~1018 findings to a broader population. 

\section{Summary} \label{sum}
We analyze 203 low-redshift changing-look AGNs (CL-AGNs; $z < 0.35$) using dual-epoch spectra from SDSS DR16 and DESI DR1. 
Spectral fitting and decomposition were performed to derive spectral parameters, 
including integrated fluxes of broad and narrow hydrogen Balmer lines, [O~{\sc iii}]~$\lambda$5007, 
continuum flux at 5100~\AA, and FWHM of broad H$\alpha$ and [O~{\sc iii}]~$\lambda$5007, 
based on the best-fit models for each epoch spectrum. 
Systematic uncertainties were quantified by incorporating the model uncertainties associated with the host-galaxy template. 

In this sample, 11.3\% of the observations are classified as Type 1.0, 26.6\% as Type 1.2, 43.1\% as Type 1.5, and 19\% as Type 1.8/2.0. 
We categorize AGN type transitions into 10 distinct cases and report their respective proportions 
across an observation time baseline ranging from 1000 to 8000 days. 
The total sample is divided into two datasets:  
Dataset A, characterized by minor type variations (110 AGNs), 
likely representing general AGN variability; 
and Dataset B, exhibiting significant type transitions (93 AGNs), 
which display the typical turn-on/turn-off behavior characteristic of CL-AGNs. 

All datasets show clear variability in the optical continuum and emission lines, 
and exhibit both “bluer-when-brighter” and “redder-when-brighter” trends. 
They reveal a robust relationship between the broad H$\beta$/[O III] $\lambda$5007 ratio 
and the broad H$\alpha$ luminosity,  
${\rm log(H\beta/[O~III])}=(0.63\pm 0.07){\rm log}(L_{\rm H\alpha})-(26.49\pm2.96)\pm0.48$ for Dataset B; 
as well as a credible relationship between H$\beta$/[O III] $\lambda$5007 and Eddington ratio, 
${\rm log(H\beta/[O~III])}=(0.59\pm 0.08){\rm log}(L_{\rm bol}/L_{\rm Edd})+(1.02\pm0.15)\pm0.53$ for Dataset B. 

The former provides direct observational support for the theoretical prediction that 
AGN type transitions across Types 1.0, 1.2, 1.5, 1.8, and 2.0 may result from changes 
in the relative ionizing flux within the broad-line region. 
Combined with the latter, 
these results suggest that variations in accretion rate modulate the ionizing flux in the broad-line region, 
thereby triggering AGN Type transitions. 
This underscores the necessity of incorporating supermassive black hole accretion processes into the AGN unification model. 
Potential causes for the substantial intrinsic scatter in these correlations have been discussed. 
Whether this scatter is linked to stellar evolution in the outer accretion disk remains an open question 
and will be explored in future work. 

\begin{acknowledgments}
We gratefully acknowledge the referees for an insightful report that enhanced the quality of the manuscript. 
This work is supported by the National Key R$\&$D Program of China with No. 2021YFA1600404. 
We acknowledge financial support from the National Natural Science Foundation of China (NSFC-12573020, 12073068, 11703077, 11873048, 11991051), 
the Young Talent Project of Yunnan Province, the Youth Innovation Promotion Association of Chinese Academy of Sciences (2022058), 
the Yunnan Province Foundation (202001AT070069), and the Light of West China Program provided by Chinese Academy of Sciences (Y7XB016001), 
the science research grants from the China Manned Space Project with No. CMS-CSST-2025-A07. 
\end{acknowledgments}


\begin{thebibliography}{99}
%
\bibitem[Antonucci(1993)]{Antonucci1993} Antonucci, R.\ 1993, \araa, 31, 473. doi:10.1146/annurev.aa.31.090193.002353
\bibitem[Bentz et al.(2013)]{Bentz2013} Bentz, M.~C., Denney, K.~D., Grier, C.~J., et al.\ 2013, \apj, 767, 149. doi:10.1088/0004-637X/767/2/149
\bibitem[Bon et al.(2018)]{Bon2018} Bon, N., Bon, E., \& Marziani, P.\ 2018, Frontiers in Astronomy and Space Sciences, 5, 3. doi:10.3389/fspas.2018.00003
\bibitem[Boroson \& Green(1992)]{Boroson1992} Boroson, T.~A. \& Green, R.~F.\ 1992, \apjs, 80, 109. doi:10.1086/191661
\bibitem[Bruzual \& Charlot(2003)]{Bruzual2003} Bruzual, G. \& Charlot, S.\ 2003, \mnras, 344, 1000. doi: 10.1046/j.1365-8711.2003.06897.x
\bibitem[Cai \& Wang(2023)]{Cai2023} Cai, Z.-Y. \& Wang, J.-X.\ 2023, Nature Astronomy, 7, 1506. doi:10.1038/s41550-023-02088-5
\bibitem[Cantiello et al.(2021)]{Cantiello2021} Cantiello, M., Jermyn, A.~S., \& Lin, D.~N.~C.\ 2021, \apj, 910, 2, 94. doi:10.3847/1538-4357/abdf4f
\bibitem[Cheng \& Wang(1999)]{Cheng1999} Cheng, K.~S. \& Wang, J.-M.\ 1999, \apj, 521, 2, 502. doi:10.1086/307572
\bibitem[Chen et al.(2023)]{Chen2023} Chen, Y.-J., Bao, D.-W., Zhai, S., et al.\ 2023a, \mnras, 520, 1807. doi:10.1093/mnras/stad051
\bibitem[Cho et al.(2023)]{Cho2023} Cho, H., Woo, J.-H., Wang, S., et al.\ 2023, \apj, 953, 2, 142. doi:10.3847/1538-4357/ace1e5
\bibitem[Cohen et al.(1986)]{Cohen1986} Cohen, R.~D., Rudy, R.~J., Puetter, R.~C., et al.\ 1986, \apj, 311, 135. doi:10.1086/164758
\bibitem[Denney et al.(2014)]{Denney2014} Denney, K.~D., De Rosa, G., Croxall, K., et al.\ 2014, \apj, 796, 134. doi:10.1088/0004-637X/796/2/134
\bibitem[Dong et al.(2008)]{Dong2008} Dong, X., Wang, T., Wang, J., et al.\ 2008, \mnras, 383, 2, 581. doi:10.1111/j.1365-2966.2007.12560.x
\bibitem[Du et al.(2018a)]{Du2018a} Du, P., Brotherton, M.~S., Wang, K., et al.\ 2018a, \apj, 869, 2, 142. doi:10.3847/1538-4357/aaed2c
\bibitem[Du et al.(2014)]{Du2014} Du, P., Hu, C., Lu, K.-X., et al.\ 2014, \apj, 782, 1, 45. doi:10.1088/0004-637X/782/1/45
\bibitem[Du et al.(2015)]{Du2015} Du, P., Hu, C., Lu, K.-X., et al.\ 2015, \apj, 806, 1, 22. doi:10.1088/0004-637X/806/1/22
\bibitem[Du et al.(2018b)]{Du2018b} Du, P., Zhang, Z.-X., Wang, K., et al.\ 2018b, \apj, 856, 1, 6. doi:10.3847/1538-4357/aaae6b
\bibitem[Feng et al.(2021)]{Feng2021} Feng, H.-C., Liu, H.~T., Bai, J.~M., et al.\ 2021, \apj, 912, 92. doi:10.3847/1538-4357/abefe0
\bibitem[Gaskell et al.(2021)]{Gaskell2021} Gaskell, C.~M., Bartel, K., Deffner, J.~N., et al.\ 2021, \mnras, 508, 4, 6077. doi:10.1093/mnras/stab2443
\bibitem[Giustini \& Proga(2019)]{Giustini2019} Giustini, M. \& Proga, D.\ 2019, \aap, 630, A94. doi:10.1051/0004-6361/201833810
\bibitem[Graham et al.(2020)]{Graham2020} Graham, M.~J., Ross, N.~P., Stern, D., et al.\ 2020, \mnras, 491, 4925. doi:10.1093/mnras/stz3244
\bibitem[Greene \& Ho(2005)]{Greene2005} Greene, J.~E. \& Ho, L.~C.\ 2005, \apj, 630, 122. doi:10.1086/431897
\bibitem[Guo et al.(2020)]{Guo2020} Guo, H., Shen, Y., He, Z., et al.\ 2020, \apj, 888, 2, 58. doi:10.3847/1538-4357/ab5db0
\bibitem[Guo et al.(2018)]{Guo2018} Guo, H., Shen, Y., \& Wang, S.\ 2018, Astrophysics Source Code Library. ascl:1809.008
\bibitem[Guo et al.(2019)]{Guo2019} Guo, H., Sun, M., Liu, X., et al.\ 2019, \apjl, 883, 2, L44. doi:10.3847/2041-8213/ab4138
\bibitem[Guo et al.(2024a)]{Guo2024a} Guo, W.-J., Zou, H., Fawcett, V.~A., et al.\ 2024a, \apjs, 270, 26. doi:10.3847/1538-4365/ad118a
\bibitem[Guo et al.(2024b)]{Guo2024b} Guo, W.-J., Zou, H., Greenwell, C.~L., et al.\ 2024b, arXiv:2408.00402. doi:10.48550/arXiv.2408.00402
\bibitem[Guo et al.(2025)]{Guo2025} Guo, W.-J., Zou, H., Greenwell, C.~L., et al.\ 2025, \apjs, 278, 1, 28. doi:10.3847/1538-4365/adc124
\bibitem[Ho(2008)]{Ho2008} Ho, L.~C.\ 2008, \araa, 46, 475. doi:10.1146/annurev.astro.45.051806.110546
\bibitem[Homan et al.(2023)]{Homan2023} Homan, D., Krumpe, M., Markowitz, A., et al.\ 2023, \aap, 672, A167. doi:10.1051/0004-6361/202245078
\bibitem[Hu et al.(2008)]{Hu2008} Hu, C., Wang, J.-M., Ho, L.~C., et al.\ 2008, \apj, 687, 1, 78. doi:10.1086/591838
\bibitem[Hu et al.(2025)]{Hu2025} Hu, C., Yao, Z.-H., Chen, Y.-J., et al.\ 2025, \apjs, 278, 2, 61. doi:10.3847/1538-4365/add40b
\bibitem[Husemann et al.(2016)]{Husemann2016} Husemann, B., Urrutia, T., Tremblay, G.~R., et al.\ 2016, \aap, 593, L9. doi:10.1051/0004-6361/201629245
\bibitem[Jana et al.(2025)]{Jana2025} Jana, A., Ricci, C., Temple, M.~J., et al.\ 2025, \aap, 693, A35. doi:10.1051/0004-6361/202451058
\bibitem[Jiang \& Pan(2025)]{Jiang2025} Jiang, N. \& Pan, Z.\ 2025, \apjl, 983, 1, L18. doi:10.3847/2041-8213/adc456
\bibitem[Kelly(2007)]{Kelly2007} Kelly, B.~C.\ 2007, \apj, 665, 2, 1489. doi:10.1086/519947
\bibitem[Kim et al.(2018)]{Kim2018} Kim, D.-C., Yoon, I., \& Evans, A.~S.\ 2018, \apj, 861, 51. doi:10.3847/1538-4357/aac77d
\bibitem[Komossa et al.(2024)]{Komossa2024} Komossa, S., Grupe, D., Marziani, P., et al.\ 2024, arXiv:2408.00089. doi:10.48550/arXiv.2408.00089
\bibitem[Korista \& Goad(2004)]{Korista2004} Korista, K.~T. \& Goad, M.~R.\ 2004, \apj, 606, 2, 749. doi:10.1086/383193 
\bibitem[Krumpe et al.(2017)]{Krumpe2017} Krumpe, M., Husemann, B., Tremblay, G.~R., et al.\ 2017, \aap, 607, L9. doi:10.1051/0004-6361/201731967
\bibitem[Lazzeroni et al.(2014)]{Lazzeroni2014} Lazzeroni, L. C., Lu, Y., \& Belitskaya-Lévy, I. \ 2014, Molecular psychiatry, 19, 12. doi: 10.1038/mp.2013.184 
\bibitem[Li et al.(2022)]{Li2022} Li, R., Ho, L.~C., Ricci, C., et al.\ 2022, \apj, 933, 70. doi:10.3847/1538-4357/ac714a
\bibitem[Liu et al.(2021)]{Liu2021} Liu, W.-J., Lira, P., Yao, S., et al.\ 2021, \apj, 915, 63. doi:10.3847/1538-4357/abf82c
\bibitem[Lu et al.(2006)]{Lu2006} Lu, H., Zhou, H., Wang, J., et al.\ 2006, \aj, 131, 2, 790. doi:10.1086/498711
\bibitem[Lu et al.(2022)]{Lu2022} Lu, K.-X., Bai, J.-M., Wang, J.-M., et al.\ 2022, \apjs, 263, 1, 10. doi:10.3847/1538-4365/ac94d3
\bibitem[Lu et al.(2025)]{Lu2025} Lu, K.-X., Li, Y.-R., Wu, Q., et al.\ 2025, \apjs, 276, 2, 51. doi:10.3847/1538-4365/ad9a5a
\bibitem[Lu et al.(2021)]{Lu2021} Lu, K.-X., Wang, J.-G., Zhang, Z.-X., et al.\ 2021, \apj, 918, 2, 50. doi:10.3847/1538-4357/ac0c78
\bibitem[Lu et al.(2019)]{Lu2019} Lu, K.-X., Zhao, Y., Bai, J.-M., et al.\ 2019, \mnras, 483, 2, 1722. doi:10.1093/mnras/sty3229
\bibitem[Lyu et al.(2021)]{Lyu2021} Lyu, B., Yan, Z., Yu, W., et al.\ 2021, \mnras, 506, 4188. doi:10.1093/mnras/stab1581
\bibitem[MacLeod et al.(2016)]{MacLeod2016} MacLeod, C.~L., Ross, N.~P., Lawrence, A., et al.\ 2016, \mnras, 457, 389. doi:10.1093/mnras/stv2997
\bibitem[Ma et al.(2025)]{Ma2025} Ma, Q.-Q., Gu, W.-M., Cai, Z.-Y., et al.\ 2025, \apj, 985, 2, 185. doi:10.3847/1538-4357/add346
\bibitem[Marziani et al.(2018)]{Marziani2018} Marziani, P., Dultzin, D., Sulentic, J.~W., et al.\ 2018, Frontiers in Astronomy and Space Sciences, 5, 6. doi:10.3389/fspas.2018.00006
\bibitem[McElroy et al.(2016)]{McElroy2016} McElroy, R.~E., Husemann, B., Croom, S.~M., et al.\ 2016, \aap, 593, L8. doi:10.1051/0004-6361/201629102
\bibitem[McLure \& Dunlop(2004)]{McLure2004} McLure, R.~J. \& Dunlop, J.~S.\ 2004, \mnras, 352, 1390. doi:10.1111/j.1365-2966.2004.08034.x
\bibitem[Mereghetti et al.(2021)]{Mereghetti2021} Mereghetti, S., Balman, S., Caballero-Garcia, M., et al.\ 2021, Experimental Astronomy, 52, 309. doi:10.1007/s10686-021-09809-6
\bibitem[Moran et al.(1996)]{Moran1996} Moran, E.~C., Halpern, J.~P., \& Helfand, D.~J.\ 1996, \apjs, 106, 341. doi:10.1086/192341
\bibitem[Nicastro \& Elvis(2000)]{Nicastro2000} Nicastro, F. \& Elvis, M.\ 2000, \nar, 44, 7-9, 569. doi:10.1016/S1387-6473(00)00101-9
\bibitem[Noda \& Done(2018)]{Noda2018} Noda, H. \& Done, C.\ 2018, \mnras, 480, 3, 3898. doi:10.1093/mnras/sty2032
\bibitem[Oknyansky et al.(2019)]{Oknyansky2019} Oknyansky, V.~L., Winkler, H., Tsygankov, S.~S., et al.\ 2019, \mnras, 483, 558. doi:10.1093/mnras/sty3133
\bibitem[Osterbrock(1977)]{Osterbrock1977} Osterbrock, D.~E.\ 1977, \apj, 215, 733. doi:10.1086/155407
\bibitem[Panda(2024)]{Panda2024FASS} Panda, S.\ 2024, Frontiers in Astronomy and Space Sciences, 11, 1479874. doi:10.3389/fspas.2024.1479874
\bibitem[Panda et al.(2022)]{Panda2022} Panda, S., Bon, E., Marziani, P., et al.\ 2022, Astronomische Nachrichten, 343, 1-2, e210091. doi:10.1002/asna.20210091
\bibitem[Panda et al.(2023)]{Panda2023} Panda, S., Bon, E., Marziani, P., et al.\ 2023, Bulletin of the Astronomical Society of Brazil, 34, 246. doi:10.48550/arXiv.2308.05831
\bibitem[Panda \& {\'S}niegowska(2024)]{Panda2024} Panda, S. \& {\'S}niegowska, M.\ 2024, \apjs, 272, 1, 13. doi:10.3847/1538-4365/ad344f
\bibitem[Planck Collaboration et al.(2020)]{Planck2020} Planck Collaboration, Aghanim, N., Akrami, Y., et al.\ 2020, \aap, 641, A6. doi:10.1051/0004-6361/201833910 
\bibitem[Pronik \& Chuvaev(1972)]{Pronik1972} Pronik, V.~I. \& Chuvaev, K.~K.\ 1972, Astrophysics, 8, 2, 112. doi:10.1007/BF01002159
\bibitem[Ricci \& Trakhtenbrot(2023)]{Ricci2023} Ricci, C. \& Trakhtenbrot, B.\ 2023, Nature Astronomy, 7, 1282. doi:10.1038/s41550-023-02108-4
\bibitem[Ruan et al.(2019)]{Ruan2019} Ruan, J.~J., Anderson, S.~F., Eracleous, M., et al.\ 2019, \apj, 883, 1, 76. doi:10.3847/1538-4357/ab3c1a
\bibitem[Saha et al.(2025)]{Saha2025} Saha, T., Krumpe, M., Markowitz, A., et al.\ 2025, \aap, 699, A205. doi:10.1051/0004-6361/202554707
\bibitem[Schlegel et al.(1998)]{Schlegel1998} Schlegel, D.~J., Finkbeiner, D.~P., \& Davis, M.\ 1998, \apj, 500, 525. doi: 10.1086/305772 
\bibitem[Shapovalova et al.(2010)]{Shapovalova2010} Shapovalova, A.~I., Popovi{\'c}, L. {\v{C}}., Burenkov, A.~N., et al.\ 2010, \aap, 509, A106. doi:10.1051/0004-6361/200912311
\bibitem[Shappee et al.(2014)]{Shappee2014} Shappee, B.~J., Prieto, J.~L., Grupe, D., et al.\ 2014, \apj, 788, 48. doi:10.1088/0004-637X/788/1/48
\bibitem[Sheng et al.(2017)]{Sheng2017} Sheng, Z., Wang, T., Jiang, N., et al.\ 2017, \apjl, 846, L7. doi:10.3847/2041-8213/aa85de
\bibitem[Shen et al.(2011)]{Shen2011} Shen, Y., Richards, G.~T., Strauss, M.~A., et al.\ 2011, \apjs, 194, 2, 45. doi:10.1088/0067-0049/194/2/45
\bibitem[Sniegowska et al.(2020)]{Sniegowska2020} Sniegowska, M., Czerny, B., Bon, E., et al.\ 2020, \aap, 641, A167. doi:10.1051/0004-6361/202038575
\bibitem[Stern \& Laor(2012a)]{Stern2012a} Stern, J. \& Laor, A.\ 2012a, \mnras, 423, 1, 600. doi:10.1111/j.1365-2966.2012.20901.x
\bibitem[Stern \& Laor(2012b)]{Stern2012b} Stern, J. \& Laor, A.\ 2012b, \mnras, 426, 4, 2703. doi:10.1111/j.1365-2966.2012.21772.x
\bibitem[Stern \& Laor(2013)]{Stern2013} Stern, J. \& Laor, A.\ 2013, \mnras, 431, 1, 836. doi:10.1093/mnras/stt211
\bibitem[Sulentic et al.(2000)]{Sulentic2000} Sulentic, J.~W., Zwitter, T., Marziani, P., et al.\ 2000, \apjl, 536, 1, L5. doi:10.1086/312717 
\bibitem[Sun et al.(2018)]{Sun2018} Sun, M., Xue, Y., Richards, G.~T., et al.\ 2018, \apj, 854, 2, 128. doi:10.3847/1538-4357/aaa890
\bibitem[Urry \& Padovani(1995)]{Urry1995} Urry, C.~M. \& Padovani, P.\ 1995, \pasp, 107, 803. doi:10.1086/133630
\bibitem[Veronese et al.(2024)]{Veronese2024} Veronese, S., Vignali, C., Severgnini, P., et al.\ 2024, \aap, 683, A131. doi:10.1051/0004-6361/202348098
\bibitem[Wang et al.(2021)]{Wang2021} Wang, J.-M., Liu, J.-R., Ho, L.~C., et al.\ 2021, \apjl, 911, 1, L14. doi:10.3847/2041-8213/abee81
\bibitem[Wang et al.(2023)]{Wang2023} Wang, J.-M., Zhai, S., Li, Y.-R., et al.\ 2023, \apj, 954, 1, 84. doi:10.3847/1538-4357/acdf48
\bibitem[Wang et al.(2024)]{Wang2024} Wang, J., Xu, D.~W., Cao, X., et al.\ 2024, \apj, 970, 85. doi:10.3847/1538-4357/ad4d89 
\bibitem[Winkler(1992)]{Winkler1992} Winkler, H.\ 1992, \mnras, 257, 677. doi:10.1093/mnras/257.4.677
\bibitem[Woo et al.(2024)]{Woo2024} Woo, J.-H., Wang, S., Rakshit, S., et al.\ 2024, \apj, 962, 1, 67. doi:10.3847/1538-4357/ad132f
\bibitem[Wu et al.(2023)]{Wu2023} Wu, J., Wu, Q., Xue, H., et al.\ 2023, \apj, 950, 106. doi:10.3847/1538-4357/acce9e
\bibitem[Wu et al.(2004)]{Wu2004} Wu, X.-B., Wang, R., Kong, M.~Z., et al.\ 2004, \aap, 424, 793. doi:10.1051/0004-6361:20035845
\bibitem[Xu et al.(2024)]{Xu2024} Xu, D.~W., Komossa, S., Grupe, D., et al.\ 2024, Universe, 10, 2, 61. doi:10.3390/universe10020061
\bibitem[Yang et al.(2018)]{Yang2018} Yang, Q., Wu, X.-B., Fan, X., et al.\ 2018, \apj, 862, 109. doi:10.3847/1538-4357/aaca3a
\bibitem[Zeltyn et al.(2024)]{Zeltyn2024} Zeltyn, G., Trakhtenbrot, B., Eracleous, M., et al.\ 2024, \apj, 966, 85. doi:10.3847/1538-4357/ad2f30
\bibitem[Zhang et al.(2022)]{Zhang2022} Zhang, W.~J., Shu, X.~W., Sheng, Z.~F., et al.\ 2022, \aap, 660, A119. doi:10.1051/0004-6361/202142253
%
\end{thebibliography}
\end{document}